\documentclass[prd,12pt,nofootinbib,preprint]{revtex4}
\usepackage[T1]{fontenc}
\usepackage[utf8x]{inputenc}
\usepackage{amsmath,amssymb}
\usepackage{epsfig}
\usepackage{url}
\usepackage{subfigure}
\usepackage{graphicx}
\usepackage{grffile}
\usepackage[usenames,dvipsnames]{color}
\usepackage{slashed}
\usepackage{array}
\usepackage{multirow}
\usepackage[colorlinks,citecolor=blue]{hyperref}
\usepackage{color}
\usepackage{cancel}
\usepackage{gensymb}

\def\a {\alpha}

\def\l {\lambda}

\def\bar {\overline}

\def\be {\begin{equation}}
\def\ee {\end{equation}}
\def\beq {\begin{equation}}
\def\eeq {\end{equation}}
\def\bea {\begin{eqnarray}}
\def\eea {\end{eqnarray}}

\newcommand{\besub}{\begin{subequations}}
\newcommand{\eesub}{\end{subequations}}

\def\beq{\begin{equation}}
\def\eeq{\end{equation}}
\def\barr{\begin{array}}
\def\earr{\end{array}}

\def\a{\alpha}

\def\l{\lambda}

\def\q2 {q^2}

\def\bt{\begin{table}}
\def\et{\end{table}}

\def\mET{E_T \hspace{-1.0em}/\;\:}

\begin{document}

\begin{flushright}
HRI-RECAPP-2021-003
\end{flushright}

\vskip 30pt 

\begin{center}
	{\Large \bf Dark Matter and Collider Searches in $S_3$-Symmetric 2HDM with Vector Like Leptons } \\
		\vspace*{1cm} {\sf ~Indrani Chakraborty$^{a,}$\footnote{indranic@iitk.ac.in},
		            ~Dilip Kumar
                  Ghosh$^{b,}$\footnote{tpdkg@iacs.res.in}, ~Nivedita
                  Ghosh$^{c,}$\footnote{niveditaghosh@hri.res.in}, 
                  ~Santosh Kumar Rai$^{c,}$\footnote{skrai@hri.res.in}}\\
		\vspace{10pt} {\small } { $^a$Department of Physics, Indian Institute of Technology, Kanpur 208 016, India
		     \\$^b$School Of Physical Sciences, Indian Association for the
                  Cultivation of Science,\\ 2A $\&$ 2B, Raja
                  S.C. Mullick Road, Kolkata 700032, India \\ $^c$
                  Regional Centre for Accelerator-based Particle
                  Physics \\ Harish-Chandra Research Institute, HBNI, 
Chhatnag Road, Jhunsi\\
Prayagraj (Allahabad) 211 019, 
India}
\end{center}

\begin{abstract}
We study the $S_3$-symmetric two Higgs doublet model by adding two generations of vector like leptons (VLL) 
which are odd under a discrete $Z_2$-symmetry. The lightest neutral component of the VLL acts as a dark 
matter (DM) whereas the full VLL set belongs to a dark sector with no mixings allowed with the standard model 
fermions. We analyse the model in light of dark matter and collider searches. We show that the DM is compatible with the current relic density data as well as satisfying all direct 
and indirect dark matter search constraints.  We choose some representative points in the model parameter space allowed by all aforementioned 
dark matter constraints and present a detailed collider analysis of multi-lepton signals {\it viz.} the mono-lepton, 
di-lepton, tri-lepton and four-lepton along with missing transverse energy in the final state using both the 
cut-based analysis and multivariate analysis respectively at the high luminosity 14 TeV LHC run.

\end{abstract}
\maketitle


\renewcommand{\thesection}{\Roman{section}}  
\setcounter{footnote}{0}  
\renewcommand{\thefootnote}{\arabic{footnote}}  


\section{Introduction}
\label{intro}

The Standard Model (SM) particle spectrum appears to be complete after the Higgs 
discovery \cite{Aad:2012tfa,Chatrchyan:2012ufa}, albeit the several shortcomings, both theoretical 
and experimental. The resolution to any of these shortcomings beg for an extension of the SM framework, 
which invariably lead us to scenarios with additional bosonic or fermionic degrees of freedom in the particle 
spectrum. One such possible scenario is an extension with vector like fermions, whose left- and right-chiral 
counterparts transform identically under the SM gauge symmetry. These VLLs can appear naturally in grand unified 
theories \cite{Rosner:2000rd,Das:2016xuc,Joglekar:2016yap,Das:2017fjf}, theories with non-minimal 
supersymmetric extensions \cite{Moroi:1991mg,Moroi:1992zk,Martin:2009bg,Babu:2008ge,Martin:2010dc,Graham:2009gy,Kang:2007ib}, warped or universal 
extra-dimension \cite{Randall:1999ee,Randall:1999vf,Agashe:2004cp,Agashe:2006wa,Li:2012zy,Huang:2012kz,Biggio:2003kp,Kaplan:2000av,Cheng:1999fw}, 
composite Higgs model \cite{Chivukula:1998wd,Dobrescu:1997nm,He:2001fz,Contino:2006qr,Anastasiou:2009rv,Kong:2011aa,Gillioz:2012se}, 
little Higgs model \cite{ArkaniHamed:2002qy,Perelstein:2003wd,Carena:2006jx,Matsumoto:2008fq,Han:2003wu}, etc. 
VLLs can acquire masses from the gauge-invariant bilinear dimension-3 bare mass terms. Since they do not achieve masses from the Yukawa couplings alone, unlike  fourth generation of chiral fermions, the vector like fermions are weakly constrained from electroweak 
precision observables and Higgs data \cite{Bahrami:2015mwa}. Although the vector like 
quarks participate in both the production and decay of Higgs boson, only the decay of Higgs boson is modified 
in presence of vector like leptons (VLLs). The rate of the process where the Higgs decays to two photons receives 
additional contribution from the charged VLLs in loop and gets modified \cite{Das:2015enc}. SM extended with 
one or more generations of VLLs have been studied earlier in \cite{Bizot:2015zaa,Beser:2016hxn, Kumar:2015tna}. The extensions 
of Higgs singlet \cite{Bell:2019mbn}, 
Higgs doublet \cite{Osoba:2012pc,Garg:2013rba,Angelescu:2016mhl,Angelescu:2015uiz}, 
Higgs triplet \cite{Bahrami:2013bsa,Bahrami:2015mwa,Bahrami:2015rqa}, 
left right symmetric model \cite{Chakdar:2013tca,Bahrami:2016has,Patra:2017gak} along with VLLs have 
been very popular in literature for dark matter (DM) and collider searches throughout.

In this work, we  study the $S_3$-symmetric two Higgs doublet model (2HDM) \cite{Das:2017zrm,Cogollo:2016dsd} 
augmented with two generations of VLLs. The need to add two generations of VLL instead of one, is to maintain 
an exact $S_3$-symmetry in the Yukawa sector. One of the primary motivations of the $S_3$-symmetric 2HDM 
is aimed at  understanding the fermion mass hierarchy within the SM, as it provides a proper description of the 
mass hierarchy and mixing among the quarks. Non-zero quark masses and non-block-diagonal CKM matrix, 
compatible with the experiment, makes this special kind of 2HDM endowed with non-abelian $S_3$ group very 
attractive. One also notes that unlike the general 2HDM, $S_3$-symmetric 2HDM naturally provides a 125 GeV 
SM-like Higgs boson, which we discuss later. As we want to study the DM phenomenology of the VLL in this 
model, we impose an additional $Z_2$ symmetry in the model under which all the VLLs are odd while the SM 
particles along with the 2HDM is even. In Ref. \cite{Angelescu:2016mhl}, the authors have used a 
$CP$-conserving 2HDM along with one generation of VLLs to study the DM phenomenology. The difference 
in their study with ours lies not only in the particle spectrum due to the presence of an extra generation of VLL 
which is mandated by the $S_3$-symmetry, but also in the way these VLLs interact with the visible sector 
particles. To be more specific, the quartic part of the Yukawa Lagrangian being $S_3$-symmetric, there exist additional interactions with respect to the reference \cite{Angelescu:2016mhl} due to the presence of an extra generation of VLLs. In addition, in our framework,
 $S_3$-symmetry is softly broken by the dimension-3 Dirac as well as Majorana mass terms unlike reference \cite{Angelescu:2016mhl}, where the authors considered only Dirac mass terms for soft breaking. In our framework the lightest neutral mass eigenstate of the VLL is a viable DM candidate, which 
produces the correct relic density, and its direct detection cross-sections and thermally averaged annihilation 
cross-sections in indirect detection are compatible with that of the experiments. 

We choose some representative benchmark points from the multi-dimensional parameter space which satisfy 
the relic density, direct and indirect search constraints and perform collider analysis for some specific multi-lepton 
channels containing mono-lepton, di-lepton, tri-lepton and four leptons along with missing transverse energy in 
the final state. Multi-lepton signals have already been analysed in the context of 
additional VLLs \cite{Freitas:2020ttd,Bissmann:2020lge}. In addition, there already exists several searches by ATLAS
and CMS, in the context of a few beyond SM models, for the final states comprising of  
mono-lepton \cite{ATLAS:2014wra,Khachatryan:2014tva,Sirunyan:2018mpc}, 
di-lepton \cite{Aad:2019vnb}, tri-lepton \cite{Aad:2019vvi,ATLAS:2020ckz} and four leptons along with 
missing transverse energy \cite{Aaboud:2018zeb,Aad:2021lzu}. We have considered the limits arising out 
of these studies in our work and highlight how the signals differ in each of our individual cases and the necessity 
to modify the cuts to optimise our signal events over the SM background. 

The paper is structured as follows. In section \ref{model} we briefly discuss the necessary extensions over the 
$S_3$-symmetric 2HDM  \cite{Das:2017zrm,Cogollo:2016dsd} done in our model. Section \ref{constraints} deals 
with the relevant theoretical and experimental constraints to be considered which is followed by the DM 
phenomenology in section \ref{DM}. Then we move on to section \ref{sec:collider} where we present the collider 
analysis of the model in the leptonic channels, namely the signals having mono-lepton, di-lepton, tri-lepton and 
four-lepton along with missing transverse energy in the final state. Finally we summarise and conclude in 
section \ref{conc}.  

\section{Model}
\label{model}
 We consider the $S_3$-symmetric 2HDM augmented with two generations of VLL. The reason for adding 
 two generations of VLL is to ensure an $S_3$-symmetric Yukawa Lagrangian. Each generation of VLL consists 
 of one left-handed lepton doublet $L_{L_i}'$, one right-handed charged lepton singlet $e_{R_i}'$, one 
 right-handed singlet neutrino $\nu_{R_i}'$ and their mirror counter parts with opposite chirality but same 
 gauge charges, {\em i.e.} $L_{R_i}'', e_{L_i}''$ and $\nu_{L_i}''$ with $i=2.$ These two generations of VLLs 
 are doublets under $S_3$-symmetry. Different quantum numbers associated with the particles are shown 
 in Table \ref{quantum} and Table \ref{quantum-S3}. In Table \ref{quantum}, $Q_{iL}, L_{iL}$ are left-handed 
 quark and lepton doublets respectively in SM with $i=1,2,3$. $u_{iR}, d_{iR}$ are right-handed up-type and 
 down-type quark singlets respectively with $i=1,2,3$.
\begin{table}[htpb!]
\begin{center}
\begin{tabular}{|c|c|c|c|c|}
\hline
\hspace{7mm} Fields \hspace{7mm} &  \hspace{5mm} $SU(2)_L$ \hspace{5mm} & \hspace{5mm} $SU(3)_C$ \hspace{5mm} & \hspace{5mm} $Y$~~ \hspace{5mm} & \hspace{5mm} $Z_2$ ~~\hspace{5mm}\\
\hline
\hline
$\phi_1$ & 2& 1& +1 & 1\\
\hline
$\phi_2$ &2 &1  & +1 & 1\\
\hline
$Q_{iL}, ~~ i =1,2,3$ &2 &3  & $+\frac{1}{3}$ &1\\
\hline
 $u_{iR}, ~~ i =1,2,3$ &1 & 3 & $+\frac{4}{3}$ & 1\\
 \hline 
 $d_{iR}, ~~ i =1,2,3$ &1 & 3 & $-\frac{2}{3}$ & 1\\
 \hline
$\ell_{iL}, ~~ i =1,2,3$ & 2& 1&  $-1$ & 1 \\
\hline
$e_{iR},  ~~ i =1,2,3$ & 1& 1 & $-2$ &  1\\ 
\hline
$L_{L_i}',~~ i =1,2$ &2 & 1& -1 & -1\\
\hline
$L_{R_i}'',~~ i =1,2$ &2 & 1& -1 & -1 \\
\hline
$e_{R_i}', ~~ i =1,2$ &1 & 1& -2 & -1\\
\hline
$e_{L_i}'',~~ i =1,2$ &1 & 1& -2 & -1\\
\hline
$\nu_{R_i}',~~ i =1,2$ &1 & 1& 0 & -1\\
\hline
$\nu_{L_i}'', ~~ i =1,2$ &1 & 1& 0 & -1\\
\hline
\end{tabular}
\end{center}
\caption{$SU(2)_L \times SU(3)_C \times U(1)_Y \times Z_2$ quantum numbers assigned to the particles in the model.}
\label{quantum}
\end{table}

%
\begin{table}[htpb!]
\begin{center}
\begin{tabular}{|c||c|}
\hline
Particles  &   $S_3$  \\ \hline \hline
$\begin{pmatrix}
\phi_1 \\
\phi_2
\end{pmatrix}$;  \hspace{2mm} 
$\begin{pmatrix}
Q_{1L} \\
Q_{2L}
\end{pmatrix}$;  \hspace{2mm} 
$\begin{pmatrix}
u_{1R} \\
u_{2R}
\end{pmatrix}$;  \hspace{2mm}
$\begin{pmatrix}
d_{1R} \\
d_{2R}
\end{pmatrix}$;  \hspace{2mm}
$\begin{pmatrix}
\ell_{1L} \\
\ell_{2L}
\end{pmatrix}$; \hspace{2mm}
$\begin{pmatrix}
e_{1R} \\
e_{2R}
\end{pmatrix}$
& 2  \\ \hline
$Q_{3L}$, $u_{3R}$, $d_{3R}$, $\ell_{3L}$, $e_{3R}$ & 1 \\ \hline \hline 
$\begin{pmatrix}
L_{L_1}' \\
L_{L_2}' \\
\end{pmatrix}$;  \hspace{5mm}
$\begin{pmatrix}
L_{R_1}'' \\
L_{R_2}'' \\
\end{pmatrix}$;  \hspace{5mm}
$\begin{pmatrix}
e_{R_1}' \\
e_{R_2}' \\
\end{pmatrix}$;  \hspace{5mm}
$\begin{pmatrix}
e_{L_1}'' \\
e_{L_2}'' \\
\end{pmatrix}$; \hspace{5mm}
$\begin{pmatrix}
\nu_{R_1}' \\
\nu_{R_2}' \\
\end{pmatrix}$;   \hspace{5mm}
$\begin{pmatrix}
\nu_{L_1}'' \\
\nu_{L_2}'' \\
\end{pmatrix}$  \hspace{5mm} & 2 \\
\hline 
\hline
\end{tabular}
\end{center}
\caption{$S_3$ quantum number assigned to the particles in the model.}
\label{quantum-S3}
\end{table}


\subsection{Scalar Lagrangian}
In the $S_3$-symmetric 2HDM, there are two $SU(2)_L$ doublets $\phi_1$ and $\phi_2$ with hypercharge 
$Y= + 1$ \footnote{The hyper-charge $Y$ has been computed by using the relation : $Q = T_3 + \frac{Y}{2}$. 
Here $T_3$ and $Q$ are the weak isospin and electric charge respectively.}, which collectively behave like a 
doublet under $S_3$-symmetry, {\em i.e.}
$\begin{pmatrix}
\phi_1 \\
\phi_2
\end{pmatrix} = \Phi$. 
For this specific doublet representation, the elements of $S_3$ is given by \cite{Das:2014fea},
\bea
\begin{pmatrix}
{\rm cos} \psi~~ {\rm sin} \psi \\
-{\rm sin} \psi~~ {\rm cos} \psi
\end{pmatrix} , ~~
\begin{pmatrix}
{\rm cos} \psi~~ {\rm sin} \psi \\
{\rm sin} \psi~~ -{\rm cos} \psi
\end{pmatrix}, {\rm for}~~ ( \psi = 0, \pm \frac{2 \pi}{3})
\eea
After symmetry breaking, $\phi_i$ can be expressed as,
\begin{eqnarray}
\phi_i = \begin{pmatrix}
\phi_i^+ \\
\frac{1}{\sqrt{2}} (v_i + h_i + i \rho_i)
\end{pmatrix}
\end{eqnarray}
Here $v_i$'s are vacuum expectation values (VEV) and $v_1 = v \cos \beta , ~v_2 = v \sin \beta , 
v = \sqrt{v_1^2 + v_2^2} = 246$ GeV. $\tan \beta$ can be defined as the ratio of two vacuum expectation 
values : $\tan \beta = \frac{v_2}{v_1}$.

The quartic part of the most general renormalisable $S_3$-symmetric scalar potential is given by \cite{Das:2017zrm},
\begin{eqnarray}
 V_4(\phi_1, \phi_2)&=&   \lambda_1 (\phi_1^\dagger\phi_1+\phi_2^\dagger\phi_2)^2  +\lambda_2 (\phi_1^\dagger\phi_2 -\phi_2^\dagger\phi_1)^2 \nonumber \\
 && + \lambda_3
\left\{(\phi_1^\dagger\phi_2+\phi_2^\dagger\phi_1)^2
+(\phi_1^\dagger\phi_1-\phi_2^\dagger\phi_2)^2\right\} \,.
\label{V4}
\end{eqnarray}
The most general quadratic part of the scalar potential is  \cite{Das:2017zrm} :
\bea
V_2(\phi_1, \phi_2) = m_{11}^2 (\phi_1^\dagger\phi_1) + m_{22}^2 (\phi_2^\dagger\phi_2) - \{m_{12}^2 (\phi_1^\dagger\phi_2) + {\rm h.c.}\}
\label{V2}
\eea

In Eq.(\ref{V4}), the quartic couplings $\lambda_1, \lambda_2$ and $\lambda_3$ are real owing to the 
hermiticity of the scalar potential. In the quadratic part of the potential in Eq.(\ref{V2}), $m_{11}^2, m_{22}^2$ are 
real, $m_{12}^2$ can be complex in principle. Throughout the analysis, we have taken $m_{12}^2$ to be 
real to avoid $CP$-violation. Though $m_{11}^2 = m_{22}^2$ and $m_{12}^2 =$ 0 ensure that the quadratic 
part of the potential is $S_3$-symmetric, this configuration ends up with a massless scalar \cite{Das:2017zrm}. 
Thus for our analysis, we stick to the configuration $m_{11}^2 = m_{22}^2$ and $m_{12}^2 \neq$ 0 (which breaks 
the $S_3$-symmetry softly),  which fixes the value of $\tan \beta =1$, following the minimisation condition of the  
scalar potential:\footnote{In Ref.~\cite{Chakrabarty:2018yoy}, it has been shown that 
with the configuration $m_{11}^2 = m_{22}^2$ and $m_{12}^2 \neq$ 0, one can still achieve the correct mass hierarchy 
in the quark sector by computing the correction to the eigenvalues using first order (non-degenerate) perturbation theory.} 
\bea
&& m_{11}^2 = m_{12}^2 \tan \beta - (\lambda_1 + \lambda_3)v^2 \,, \nonumber \\
&& m_{22}^2 = m_{12}^2 \cot \beta - (\lambda_1 + \lambda_3)v^2
\eea
The scalar sector of this model consists of SM-like Higgs ($h$), heavy Higgs ($H$), pseudoscalar Higgs ($A$) 
and charged Higgs ($H^{\pm}$). The limit at which $h$ behaves as SM-like Higgs boson is defined as 
the {\em alignment limit}. This limit is naturally achieved in this model \cite{Das:2017zrm}.

Transformations from flavour basis to mass basis of the scalars occur through the following $ 2 \times 2 $ 
orthogonal matrix :

\begin{eqnarray}
\begin{pmatrix}
w^\pm (z) \\
H^\pm (A)
\end{pmatrix} = \begin{pmatrix}
                \cos \beta & \sin \beta \\
                -\sin \beta & \cos \beta
                \end{pmatrix}
                \begin{pmatrix}
                \phi_1^\pm (\rho_1) \\
                \phi_2^\pm (\rho_2)
                \end{pmatrix}
\end{eqnarray}
$w^\pm$ and $z$ being the charged and neutral Golstone boson respectively.
 
The light Higgs and the heavy Higgs of the model are connected to there flavour eigenstates via,
\begin{eqnarray}
\begin{pmatrix}
h \\
H
\end{pmatrix} = \begin{pmatrix}
                \cos \beta & \sin \beta \\
                -\sin \beta & \cos \beta
                \end{pmatrix}
                \begin{pmatrix}
                h_1 \\
                h_2
                \end{pmatrix}
\end{eqnarray}

In Eq.(\ref{V4}), the quartic couplings $\lambda_1, \lambda_2, \lambda_3$ can be expressed in terms of 
the physical scalar masses as :

\begin{eqnarray}
\lambda_1 &=& \frac{M_h^2 - M_H^2 + M_{H^\pm}^2}{2 v^2}  \,, \nonumber \\
\lambda_2 &=& \frac{(M_{H^\pm}^2 - M_A^2)}{2 v^2}  \,, \nonumber \\ 
\lambda_3 &=& \frac{(M_H^2 - M_{H^\pm}^2)}{2 v^2} \,,
\end{eqnarray}

\subsection{Yukawa Lagrangian} 
 
The dimension-4 terms in $S_3$-symmetric Yukawa Lagrangian involving two generations of VLLs can be written as,
\begin{eqnarray}
\mathcal{L}_4 &=& -y_2[(\bar{L_{L_1}'}\tilde{\phi_2}+\bar{L_{L_2}'}\tilde{\phi_1})\nu_{R_{1}}' +  
(\bar{L_{L_1}'}\tilde{\phi_1}-\bar{L_{L_2}'}\tilde{\phi_2})\nu_{R_{2}}']
-y_4[(\bar{L_{R_1}''}\tilde{\phi_2}+\bar{L_{R_2}''}\tilde{\phi_1})\nu_{L_{1}}'' \nonumber \\
&& + (\bar{L_{R_1}''}\tilde{\phi_1}-\bar{L_{R_2}''}\tilde{\phi_2})\nu_{L_{2}}'']
- y_2'[(\bar{L_{L_1}'}\phi_2+\bar{L_{L_2}'}\phi_1)e_{R_{1}}'+
(\bar{L_{L_1}'}\phi_1 - \bar{L_{L_2}'}\phi_2)e_{R_{2}}'] \nonumber \\
&& - y_4' [ (\bar{L_{R_1}''} \phi_2 + \bar{L_{R_2}''} \phi_1) e_{L_{1}}'' +
(\bar{L_{R_1}''}\phi_1-\bar{L_{R_2}''}\phi_2)e_{L_{2}}'']
+ {\rm h.c.}
\label{dim4}
\end{eqnarray}

Next we write down the dimension-3 Dirac and Majorana mass terms present in the Yukawa Lagrangian, which break $S_3$-symmetry 
softly:\footnote{Note that with exact $S_3$-symmetry, the mass of the vector like leptons will be proportional to the product 
of Yukawa coupling and the electroweak vacuum expectation value (VEV), which in turn will lead to non-perturbative Yukawa 
couplings (for vector lepton masses $\sim$ 1 TeV). With $S_3$ softly broken we can write down gauge-invariant bilinear 
dimension-3 interaction terms in the Lagrangian.}

\begin{eqnarray}
\mathcal{L}_{3} &&= - M_{1}\, \bar{L_{L_1}'} \, L_{R_1}''
- M_{2}\, \bar{L_{L_1}'} \, L_{R_2}'' - M_{3}\, \bar{L_{L_2}'} \, L_{R_1}'' 
- M_{4}\, \bar{L_{L_2}'} \, L_{R_2}'' 
- \frac{1}{2} M_{5} \, \bar{\nu_{L_{1}}^{c ~ ''}} \, \nu_{L_{1}}'' 
- \frac{1}{2} M_{6} \, \bar{\nu_{L_{2}}^{c~''}} \, \nu_{L_{2}}'' \nonumber \\
&& - \frac{1}{2} M_{7} \, \bar{\nu_{R_{1}}^{c~'}} \, \nu_{R_{1}}'
- \frac{1}{2} M_{8} \, \bar{\nu_{R_{2}}^{c~'}} \, \nu_{R_{2}}' 
- M_{9} \, \bar{\nu_{L_{1}}''} \, \nu_{R_{1}}'  - M_{10} \, \bar{\nu_{L_{1}}''} \, \nu_{R_{2}}' 
 - M_{11} \, \bar{\nu_{L_{2}}''} \, \nu_{R_{1}}' - M_{12} \, \bar{\nu_{L_{1}}''} \, \nu_{R_{2}}'  \nonumber \\
&& - M_{L_{1}} \, \bar{e_{L_{1}}''} \, e_{R_{1}}' - M_{L_{2}} \, \bar{e_{L_{2}}''} \, e_{R_{2}}' 
- M_{L_{3}} \, \bar{e_{L_{1}}''} \, e_{R_{2}}' - M_{L_{4}} \, \bar{e_{L_{2}}''} \, e_{R_{1}}'  \, +  \, {\rm h.c.}
\label{majorana}
\end{eqnarray}
Here the fields with superscript "$c$" are the charge conjugated fields. The subscripts of "$\mathcal{L}$" in Eq.(\ref{dim4}) and Eq.(\ref{majorana}) denote the mass dimensions of the operators. Thus whole Yukawa lagrangian can be written as the sum of $\mathcal{L}_3$ and $\mathcal{L}_4$ as :
\bea
\mathcal{L}_{\rm Yuk} = \mathcal{L}_3 + \mathcal{L}_4 \,.
\eea

Here we can construct eight neutral mass eigenstates ($N_i , i = 1$..8) out of two generations of vector leptons. 
To ensure that the lightest VLL $N_1$ is the DM candidate, we impose a $Z_2$-symmetry (mentioned in 
Table \ref{quantum}) under which all the VLLs are odd and all the SM leptons 
are even. This forbids the mixing between the SM leptons and VLLs \footnote{The couplings of the SM fermions with the scalar doublets $\phi_1$ and $\phi_2$ are also governed by $S_3$-symmetry. The most general $S_3$-symmetric Yukawa lagrangian containing dimension-4 interactions between the SM fermions and the scalar doublets can be found in \cite{Das:2017zrm}.}. The two Higgs doublets are assumed to 
be even under this $Z_2$-symmetry. 


In this set up, $8 \times 8 $ neutral VLL mass matrix in the basis $(\nu_{L_1}^{'c}, \nu_{R_1}', \nu_{R_1}'', 
\nu_{L_1}^{''c}, \nu_{L_2}^{'c}, \nu_{R_2}', \nu_{R_2}'', \nu_{L_2}^{''c})^T$  reads :
\bea
&&\frac{1}{2} \left(~~\bar{\nu_{L_1}'} ~~ \bar{\nu_{R_1}^{'c}} ~~ \bar{\nu_{R_1}^{''c}} ~~ \bar{\nu_{L_1}''} ~~ \bar{\nu_{L_2}'} ~~ \bar{\nu_{R_2}^{'c}} ~~ \bar{\nu_{R_2}^{''c}} ~~ \bar{\nu_{L_2}''} ~~\right) M_\nu \begin{pmatrix}
\nu_{L_1}^{'c} \\
\nu_{R_1}' \\
\nu_{R_1}'' \\
\nu_{L_1}^{''c} \\
\nu_{L_2}^{'c} \\
\nu_{R_2}' \\
\nu_{R_2}'' \\
\nu_{L_2}^{''c}
\end{pmatrix} \nonumber 
\eea
With 
\bea
M_\nu = \begin{pmatrix} 
0 & \frac{y_2 v_2}{\sqrt{2}} & M_1 & 0 & 0 & \frac{y_2 v_1}{\sqrt{2}} & M_2 & 0 \\
\frac{y_2 v_2}{\sqrt{2}} & M_7 & 0 & M_9 & \frac{y_2 v_1}{\sqrt{2}} & 0 & 0 & M_{11} \\
M_1 & 0 & 0 & \frac{y_4 v_2}{\sqrt{2}} & M_3 & 0 & 0 & \frac{y_4 v_1}{\sqrt{2}} \\
0 & M_9 & \frac{y_4 v_2}{\sqrt{2}} & M_5 & 0 & M_{10} & \frac{y_4 v_1}{\sqrt{2}} & 0 \\
0 & \frac{y_2 v_1}{\sqrt{2}} & M_3 & 0 & 0 & \frac{ - y_2 v_2}{\sqrt{2}} & M_4 & 0 \\
\frac{y_2 v_2}{\sqrt{2}} & 0 & 0 & M_{10} & \frac{ - y_2 v_2}{\sqrt{2}} & M_8 & 0 & M_{12} \\
M_2 & 0 & 0 &\frac{y_4 v_1}{\sqrt{2}} & M_4 & 0 & 0 & \frac{ - y_4 v_2}{\sqrt{2}} \\
0 & M_{11} & \frac{y_4 v_1}{\sqrt{2}} & 0 & 0 & M_{12} & \frac{ - y_4 v_2}{\sqrt{2}} & M_6 
\end{pmatrix} \label{neutral-mat} 
\end{eqnarray}

Since $M_\nu$ is hermitian, it can be brought to diagonal form by the following transformation via unitary matrix $V$ :
\begin{equation}
V^\dag M_\nu V = {\rm diag} ~(M_{N_1}, M_{N_2}, M_{N_3}, M_{N_4}, M_{N_5}, M_{N_6}, M_{N_7}, M_{N_8}).
\end{equation}
Among all states $N_1$ is the lightest and $M_{N{j+1}} > M_{N_j}$ for $j = 1,2,...7$.

In the charged VLL sector, the mass matrix is,
\bea
&& \left(~~\bar{e_{L_1}'} ~~ \bar{e_{L_1}''} ~~ \bar{e_{L_2}'} ~~ \bar{e_{L_2}''} ~~\right) M_c \begin{pmatrix}
e_{R_1}' \\
e_{R_1}'' \\
e_{R_2}' \\
e_{R_2}'' 
\end{pmatrix} \nonumber
\eea
where,
\bea
M_c = \begin{pmatrix} 
\frac{y_2' v_2}{\sqrt{2}} & M_1 & \frac{y_2' v_1}{\sqrt{2}} & M_2  \\
M_{L_1} & \frac{y_4' v_2}{\sqrt{2}} & M_{L_3} & \frac{y_4' v_1}{\sqrt{2}} \\
\frac{y_2' v_1}{\sqrt{2}} & M_3 & -\frac{y_2' v_2}{\sqrt{2}} & M_4  \\
M_{L_4} & \frac{y_4' v_1}{\sqrt{2}} & M_{L_2} & -\frac{y_4' v_2}{\sqrt{2}}  \\ 
\end{pmatrix} \label{charged-mat}
\eea
The non-hermitian $M_c$ can be diagonalised by using the following bi-unitary transformation, with the unitary 
matrices $U_L$ and $U_R$,
\bea
U_L^\dag M_c U_R = {\rm diag} ~(M_{E_1^+}, M_{E_2^+}, M_{E_3^+}, M_{E_4^+})
\eea
Here we follow the same convention as the neutral sector, {\em i.e.} $M_{E_{i+1}^+} > M_{E_{i}^+}$ for $i= 1,2,3.$
\section{Constraints to be considered}
\label{constraints}
The $S_3$-symmetric 2HDM model has an extended scalar sector and we have included VLLs in our model with some 
being $SU(2)_L$ doublets . The addition of particles under a new symmetry which are not singlets under the SM 
gauge symmetry would lead to interactions and mixings that could affect several existing experimental observations. In 
addition, the new parameters in the model would also have to adhere to theoretical constraints to make the 
model mathematically consistent. We look at the most relevant ones and extract the constraints they could 
put on the model parameters.
\subsection{Theoretical constraints}
\begin{itemize}
\item {\bf Perturbativity}: The quartic couplings $\lambda_1. \lambda_2, \lambda_3$ are taken to be 
perturbative: $|\lambda_i| < 4 \pi,~~ i=1,2,3.$ For Yukawa couplings the corresponding bound 
reads: $|y_2|, |y_2'|, |y_4|, |y_4'| < \sqrt{4 \pi}.$
\item {\bf Stability conditions of the potential}: The quartic couplings $\lambda_1, \lambda_2$ and $\lambda_3$ are 
also constrained from the stability conditions, so that the potential remains bounded from below in any field direction:
\begin{eqnarray}
&&\lambda_1 + \lambda_3 \geq 0 \,, \nonumber \\
&&\lambda_1 \geq 0 \,, \nonumber \\
&& 2 \lambda_1 - \lambda_2 + \lambda_3 \geq 0 \,.
\end{eqnarray}
\item {\bf Higgs mass }: We keep the SM-like Higgs mass within the range: 125.1 $\pm $ 0.14 GeV \cite{Tanabashi:2018oca}.  We have used the SM-like Higgs mass as an input parameter and fix its value to $M_h = 125$ GeV throughout the analysis.
\end{itemize}

\subsection{Constraints from electroweak precision observables}

The additional extra scalars and VLLs that are not SM singlets would interact with the $W$ and $Z$ boson. 
There contributions in the self-energy correction diagrams would modify the mass of the weak gauge bosons 
and related electroweak precision observables parametrised by the oblique parameters $S, T$ and 
$U$~\cite{Peskin:1990zt,Peskin:1991sw}. Using $ M_h = 125$ GeV and top mass as 172.5 GeV, the 
allowed ranges~\cite{Haller:2018nnx} are

\begin{equation}
 \Delta S = 0.04 \pm 0.11 , \Delta T = 0.09 \pm 0.14, \Delta U = -0.02 \pm 0.11
\end{equation}


Notably the deviations in the $T$-parameter from its SM value enforces the mass splitting between the neutral and 
the charged scalar to be less than $\sim 50-100$ GeV. Regarding the contributions coming from the VLL counterpart, the differences between the Yukawa couplings $|y_2 - y_2'|$ and $|y_4 - y_4'|$ should be small to evade the bound coming from $T$-parameter \cite{Bahrami:2015mwa}.

\subsection{Higgs signal strength}
Since we demand that the lightest $CP$-even scalar $h$ is the SM like Higgs, it is imperative that we should check 
whether the production and decay of $h$ in our model is consistent with the current experimental data. The 
compatibility can be checked by computing the signal strength in the $i$th  decay mode of $h$ as, 
\begin{eqnarray}
\mu_i = \frac{\sigma^{S_3 \rm{2HDM}}(pp \rightarrow h)~ {\rm BR^{S_3 \rm{2HDM}}}(h \rightarrow i)}{\sigma^{\rm{SM}}(pp \rightarrow h)~ {\rm BR^{SM}}(h \rightarrow i)}\,.
\label{sig-str-1}
\end{eqnarray}
Assuming gluon-gluon fusion to be the most dominant Higgs production process at the LHC, one can rewrite the signal 
strength $\mu_i$ as,
\bea 
\mu_i &=& \frac{\sigma^{\rm{S_3 \rm{2HDM}}}(gg \rightarrow h)}{\sigma^{\rm{SM}}(gg \rightarrow h)} ~\frac{\Gamma_i^{\rm{S_3 \rm{2HDM}}}(h \rightarrow i)}{\Gamma_{\rm{tot}}^{\rm{S_3 \rm{2HDM}}}} ~\frac{\Gamma_{\rm{tot}}^{\rm{SM}}}{\Gamma_i^{\rm{SM}}(h \rightarrow i)} \nonumber \\
&=& \frac{\Gamma^{\rm{S_3 \rm{2HDM}}}(h \rightarrow gg)}{\Gamma^{\rm{SM}}(h \rightarrow gg)} ~\frac{\Gamma_i^{\rm{S_3 \rm{2HDM}}}(h \rightarrow i)}{\Gamma_{\rm{tot}}^{\rm{S_3 \rm{2HDM}}}} ~\frac{\Gamma_{\rm{tot}}^{\rm{SM}}}{\Gamma_i^{\rm{SM}}(h \rightarrow i)}
\eea
where $\Gamma_{\rm{tot}}$ stands for the total decay width of SM like Higgs.

Since the {\em Alignment limit} is maintained naturally in this model, the signal strengths in the decay 
channels of $h$ into $WW$ \cite{Aad:2019lpq,Sirunyan:2018egh}, $ZZ$ \cite{ATLAS:2018bsg,CMS:2019chr}, 
$b\bar{b}$ \cite{Aaboud:2018gay,CMS:2016mmc}, $\tau^+\tau^-$ \cite{ATLAS:2018lur,Sirunyan:2017khh} are 
satisfied without any loss of generality. On the other hand, the loop-induced decay mode of the  
$h \to \gamma \gamma$ will get additional contribution from the charged vector leptons and the non-standard 
charged scalars. For the chosen benchmark points, the $h \to \gamma \gamma$ signal 
strength remains within 2$\sigma$-range of the current experimental value \cite{Aaboud:2018xdt,Sirunyan:2018ouh}. 
The detailed formula for the decay width of $h \to \gamma \gamma$ channel is relegated to Appendix \ref{app : A}.

\section{Dark Matter phenomenology}
\label{DM}

As mentioned earlier, the lightest neutral VLL $N_1$ is a viable DM candidate due to its stable nature ensured 
by the imposed $Z_2$-symmetry. In this section we investigate the parameter space spanned by relevant 
and independent model parameters which are compatible with relic density \cite{Aghanim:2018eyx}, direct and 
indirect DM searches. Setting the mass of the SM-like Higgs to  125 GeV, we scan the independent 
parameters of the model in the following range :

\bea
&& |y_2| \in [1.0 : 3.0] \,, ~ |y_2'| \in [1.0 : 3.0] \,, ~ |y_4| \in [0.5 : 2.0] \,, ~ |y_4'| \in [1.0 : 3.0] \,, \nonumber \\
&& M_{L_1} \in [50 ~{\rm GeV} : 500 ~{\rm GeV}], ~~ M_{L_2} \in [50 ~{\rm GeV} : 500 ~{\rm GeV}] \,, \nonumber \\
&& M_5 \in [10 ~{\rm GeV} : 500 ~{\rm GeV}], ~ M_6 \in [10 ~{\rm GeV} : 500 ~{\rm GeV}] \,, \nonumber \\
&& M_7 \in [10 ~{\rm GeV} : 1 ~{\rm TeV}], ~ M_8 \in [10 ~{\rm GeV} : 500 ~{\rm GeV}] \,, \nonumber \\
&& M_H \in [500 ~{\rm GeV} : 800 ~{\rm GeV}], ~ M_{H^+} \in [500 ~{\rm GeV} : 800 ~{\rm GeV}] \,, \nonumber \\
&& M_A \in [500 ~{\rm GeV} : 800 ~{\rm GeV}] \,.
\eea

For the analysis, we derive the interactions, mass and mixings in the model which is then implemented in 
\texttt{FeynRules}~\cite{Alloul:2013bka}.  The \texttt{CALCHEP}~\cite{Belyaev:2012qa} model files 
obtained from \texttt{FeynRules} is made compatible to use with \texttt{ micrOMEGAs}~\cite{Belanger:2014vza}. 
The DM observables like relic density ($\Omega_{\rm DM} h^2$)\footnote{$\Omega_{\rm DM}$ is defined as the 
ratio of non-baryonic DM density to the critical density of the universe and $h$ is the reduced Hubble parameter (not 
to be confused with SM Higgs $h$).}, spin-dependent ($\sigma_{\rm SD}$) and independent 
cross-sections ($\sigma_{\rm SI}$), thermally averaged annihilation cross-sections ($ \langle \sigma v \rangle$) etc. 
are computed with the help of \texttt{ micrOMEGAs}. The observed relic abundance obtained from the PLANCK 
experiment \cite{Aghanim:2018eyx} lies in the band: $0.1166 \leq \Omega_{\rm DM} h^2 \leq 0.1206$. 
Furthermore the parameter space is also constrained by the bounds coming from several direct detection 
experiments like LUX \cite{Akerib:2016vxi}, PANDAX-II \cite{Cui:2017nnn}, Xenon-1T \cite{Aprile:2018dbl}, 
PICO \cite{Amole:2019fdf},  etc. and from the indirect detection bounds coming from 
FERMI-LAT \cite{Daylan:2014rsa}, MAGIC \cite{Ahnen:2016qkx} and PLANCK \cite{Aghanim:2018eyx} experiments. 

\begin{table}[htbp!]
\centering
	\resizebox{16cm}{!}{
	\begin{tabular}{|p{2cm}|c|c|c|c|p{3cm}|p{3cm}|}
		\hline
		Benchmark Points &  $M_{\rm DM}$  & $\Omega_{\rm DM} h^2$ & $\sigma_{\rm SD}$  & $\sigma_{\rm SI}$ &
		Annihilation cross-section  & Annihilation mode \\ 
		 &   $(\rm GeV)$ &  & (cm$^2$) &  (cm$^2$) & $\langle \sigma v \rangle$ ($cm^3/s$) &  \\ \hline
 BP1 & 81.3    & $1.04\times 10^{-1}$ & $3.4\times10^{-42} $ & $4.4\times10^{-49}$ &$2.41\times10^{-28}$ & $W^{+}W^{-}$ \\ \hline
 BP2 & 90.7 & $7.7\times10^{-2}$ & $3.6\times10^{-43} $ & $2.2\times10^{-49}$ & $5.67\times10^{-27}$ & $W^{+}W^{-}$ \\ \hline
 BP3 & 108.6  &  4.7$\times 10^{-2}$ &$1.2\times10^{-42}$ & $4.6\times10^{-49}$ & $2.72\times10^{-26}$ & $ZZ$ ( 63\%) \\
 & & & & & &  $W^{+}W^{-}$ (28\%) \\ \hline
 BP4 & 193.8    & $9.36\times 10^{-4}$ & $5.98\times10^{-42}$ & $2.54\times10^{-49}$ &$3.79\times10^{-26}$ & $ZZ$ ( 53.8\%)\\
 & & & & & &  $W^{+}W^{-}$ (46.1\%) \\ \hline
 BP5 & 282.8   & $5.21\times 10^{-4}$ & $9.07\times10^{-43}$ & $2.74\times10^{-48}$ &$2.08\times10^{-26}$ &  $ZZ$ (52.5\%) \\ 
 & & & & & &  $W^{+}W^{-}$ (47\%) \\ \hline
			\end{tabular}}
	\caption{ DM masses along with
		DM relic density, spin-dependent, spin-dependent cross-sections, thermally averaged annihilation cross-sections and dominant annihilation modes for five benchmarks.}
	\label{table:bp}
\end{table}

Having discussed what could be the possible constraints coming from the DM sector, let us illustrate it 
more in a model specific manner. The relic density can be computed as the function of the thermally averaged 
DM pair annihilation cross-sections. Since the lightest neutral VLL $N_1$ in this model is the admixture of $SU(2)$ 
singlets and doublets, it has couplings with both $W^\pm$ and $Z$-bosons. Depending on the mass of DM, 
the $s$-channel pair annihilation of the DM to $W^+ W^-, ZZ, Zh, hh, f\bar{f}$ mediated by $h, H, A$ and 
$Z$-boson can occur. Besides, $t$-channel annihilation to $ZZ,Zh,hh$ ($W^+ W^-$) via 
$N_i,~ i=1,2,...,8$ ($E_j^\pm,~ j= 1,2,..,4$) as mediator also contribute to the $ \langle \sigma v \rangle$. 

\begin{figure}[b!]
\centering
\includegraphics[width=11cm,height=8cm, angle=0]{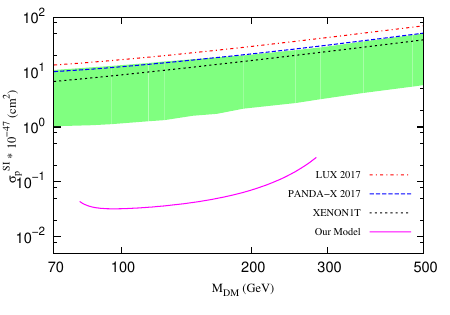} 
\caption{ Spin-independent cross-section of the proton as a function of the DM mass in our model(magenta  curve), 
and the experimental upper limits from LUX(red curve)\cite{Akerib:2016vxi}, PANDA-X(blue curve)\cite{Cui:2017nnn}, 
XENON1T (dotted black curve with 90\% C.L. and 2$\sigma$ sensitivity bands (green band))\cite{Aprile:2018dbl}.}
\label{fig:SI} 
\end{figure}
In Table \ref{table:bp}, we present five representative points BP1, BP2, BP3, BP4 and BP5 with increasing DM mass, 
which satisfy the relic density constraints as well as the direct and indirect detection bounds. Corresponding values 
of $\sigma_{\rm SI}$, $\sigma_{\rm SD}$ and $\langle \sigma v \rangle$ are also tabulated in the same table. Since 
the DM is Majorana-like, due to the $Z$-mediated process, for all the benchmarks $\sigma_{\rm SI} < \sigma_{\rm SD}$.  As mentioned earlier, from the minimisation conditions of the scalar potential of the $S_3$-symmetric 2HDM, with $m_{11}^2 = m_{22}^2$ and $m_{12}^2 \neq 0$, $\tan \beta$ is fixed to 1. Now $Hf \bar{f}$ and $A f \bar{f}$ ("$f$" is SM fermion) couplings being proportional to $(\cos \beta -\sin \beta)$, vanish at $\tan \beta = 1$ limit. Thus $s$-channel annihilations into SM fermions mediated by $H$ or $A$ are absent in this framework. The only surviving $s$-channel annihilation to SM fermions is mediated by $h$.
For the first two points,  since $M_{N_1} < M_Z$, the DM pair dominantly annihilate into $W^+ W^-$. After 
crossing the $ZZ$-threshold, the major annihilation occurs to the final state $ZZ$ along 
with $W^+ W^-$. Since the $H, A$-mediated $s$-channel annihilation to $W^+ W^-$ and $ZZ$ are forbidden at 
alignment limit, $t$-channel annihilation to $W^+ W^-$ and $ZZ$ via $E_i^\pm$ becomes the major contributor. 
Moderate $Z N_i N_j$ couplings (with $ i \neq j$) participating in the annihilation come out to be the main reason 
behind this dominance. To put this in perspective, we list the dominant annihilation modes for the aforementioned 
five benchmark points in Table \ref{table:bp} too. We note that since the {\it alignment limit} is maintained naturally 
in this model, the $s$-channel $H,A$-mediated processes leading to $W^+ W^-$ and $ZZ$ final state will not 
contribute to $ \langle \sigma v \rangle$. The scattering of DM with the nuclei within the detector material mediated 
by $Z$ or $h$, gives rise to spin-dependent and spin-independent cross-sections ($\sigma_{\rm SD}$ and 
$\sigma_{\rm SI}$) respectively, which in turn are constrained from direct detection experiments. This forces 
$h N_1 N_1$ and $Z N_1 N_1$ couplings to be small enough to circumvent the direct detection bound. This is merely a choice and the smallness of the aforementioned couplings is achieved by tuning relevant parameters of the model. Due 
to the Majorana nature of the DM, the WIMP-nucleon cross-section is dominated by spin-dependent interactions 
mediated by $Z$ boson. Hence we have to consider the direct detection bound on the $\sigma_{\rm SD}$ coming 
from the PICO experiment \cite{Amole:2019fdf}.

In Fig.\ref{fig:SI}, we depict the variation of spin-independent cross-section ($\sigma_{\rm SI}$) with DM mass 
predicted by our model (magenta curve). The black line and the green band correspond to 90\% confidence 
level (C.L.) and 2$\sigma$ sensitivity predicted by Xenon-1T experiment. We can conclude that for the dark 
matter mass range allowed by relic density constraint, $\sigma_{\rm SI}$ are smaller and allowed by the experimental 
limit shown by the black line in Fig.\ref{fig:SI}. Therefore the spin-independent cross-sections for all the chosen 
benchmarks evade the constraint coming from the direct detection experiments. As mentioned earlier, the strongest 
bound on spin-dependent cross-section comes from PICO experiment~\cite{Amole:2019fdf}. For the chosen 
benchmark points, the spin-dependent cross-section remain below the experimental limit as can be seen from 
\begin{figure}[h!]
\centering
\includegraphics[width=11cm,height=8cm, angle=0]{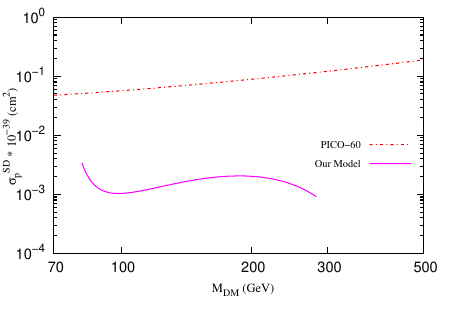} 
\caption{ Spin-dependent cross-section of proton as a function of the DM mass in our model (magenta  curve), 
and the experimental upper bound from PICO-60~\cite{Amole:2019fdf} experiment.}
\label{fig:SD} 
\end{figure}
Fig.~\ref{fig:SD}. 

\begin{figure}[htbp!]
\centering
\includegraphics[width=11cm,height=8cm, angle=0]{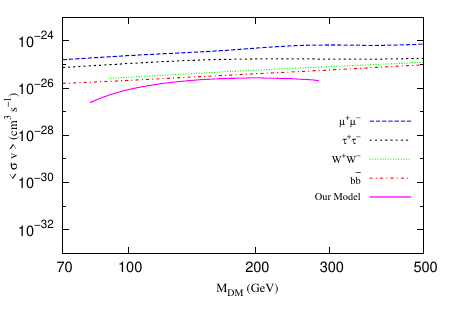} 
\caption{Variation of thermally averaged annihilation cross-section with DM masses. The magenta and the dashed 
curves represent the variations of $\langle \sigma v \rangle$ for the model and various annihilation cross-sections 
predicted by combined results of FERMI-LAT and MAGIC experiments~\cite{Ahnen:2016qkx} respectively. 
Colour coding is expressed in legends.}
\label{fig:indirect} 
\end{figure}

Indirect detection experiments look for annihilation of the DM pair to SM particles through various channels that could 
produce distinctive signatures in cosmic rays. In Fig.~\ref{fig:indirect}, we show the variation of the thermally 
averaged annihilation cross-section as a function of dark matter mass. The magenta curve signifies the variation of 
annihilation cross-section for our model. Combined results from the FERMI-LAT and MAGIC 
experiments~\cite{Ahnen:2016qkx} are represented by the dashed lines. Here the blue, black, red and green 
dashed curves show the variation of $\langle \sigma v \rangle$ with DM masses for the annihilation to 
$\mu \mu, \, \tau \tau, \, b \bar{b}$ and $W^+ W^-$ respectively. We find that the parameter space characterised 
by our benchmarks survive the bounds coming from the indirect detection experiments. We however note that 
for the DM mass range of 100-200 GeV lies quite close to the experimental bounds from the indirect detection and 
may become sensitive to future data from indirect detection experiments. We have also incorporated the experimental 
results obtained from PLANCK data \cite{Aghanim:2018eyx} in our analysis, though we have not shown it in 
Fig.\ref{fig:indirect}. We have checked that the curve representing our model in Fig.\ref{fig:indirect} lies well below the 
experimental limits from PLANCK.

\section{Collider Searches}
\label{sec:collider}

In this section we focus on the collider phenomenology of our model. We study the most likely signals of the 
model that may manifest itself at the current and future runs of the large hadron collider (LHC). As the model 
consists of an extension of the spectrum in the electroweak and leptonic sector, it becomes quite clear that the 
production of these new exotic particles would be limited by their cross section if they are too heavy. In fact, 
the limits on weakly interacting BSM particles are still quite weak 
from LHC. In this model, VLLs with unbroken $Z_2$ symmetry have no mixing with the SM fermions. Thus, the 
production of these VLL's will have to be in pairs and they would decay to a SM particle and a lighter component of 
the VLL.  We therefore focus on the relatively lighter spectrum of the exotics 
whose lightest neutral component is the DM candidate, represented by the states for the five benchmark points (BP) shown 
in Table~\ref{table:bp} and consistent with the DM phenomenology presented in the previous section. The mass of the 
remaining VLL components which correspond to the same five BP's {\it viz.} $M_{N_i}, $ with $i=1,2,...,8$ and 
$M_{E_j^\pm}$ with $j=1,2,...,4$ are tabulated in Table \ref{table:masses}.
\begin{table}[htbp!]
	\centering
	\resizebox{16cm}{!}{
	\begin{tabular}{|p{2.1cm}|c|c|c|c|c|c|c|c|c|c|c|c|c|c|}
		\hline
		Benchmark  & $M_{N_1}$ & $M_{N_2}$  & $M_{N_3}$ & $M_{N_4}$  & $M_{N_5}$ & $M_{N_6}$ & $M_{N_7}$ &  $M_{N_8}$
		&$M_{E^{\pm}_1}$ & $M_{E^{\pm}_2}$ & $M_{E^{\pm}_3}$ & $M_{E^{\pm}_4}$\\ 
	Points &$(\rm GeV)$ &$(\rm GeV)$& $(\rm GeV)$ & $(\rm GeV)$ &  $(\rm GeV)$ & $(\rm GeV)$ & $(\rm GeV)$ & $(\rm GeV)$ & $(\rm GeV)$ &
		 $(\rm GeV)$ & $(\rm GeV)$ & $(\rm GeV)$\\ \hline
		BP1  & 81.3  & 86.9 & 119.3 & 154.4 &  211.6 &   268.7 & 688.4 & 856.9 & 171.0 & 211.8 & 260.0 & 322.0 \\ \hline
		BP2  & 90.7  & 105.9 & 132.2 & 168.0  & 209.8   & 258.2 & 638.2 & 696.9 & 187.5 & 213.9 & 261.1 &  298.0\\ \hline
		BP3  & 108.3  & 119.4 & 156.8 & 188.4 & 247.4   & 249.0  & 598.9 & 648.7 & 210.7 & 262.5 & 353.6  & 452.3 \\ \hline
		BP4 & 193.8 & 204.8 & 239.8 & 245.7 & 268.3 &  274.9  & 454.5 & 494.7 & 280.2 & 313.0 & 356.8 & 398.5  \\ \hline
		BP5 & 282.8 & 323.0 &  327.7 & 333.4  & 364.7 & 375.0 & 463.1 & 483.7 & 376.8 & 388.0 & 458.5 & 472.2  \\ \hline
		
	\end{tabular}}
	\caption{Masses of neutral and charged VLLs for five benchmarks.}
	\label{table:masses}
\end{table}

The pair production of the VLL would give rise to lepton rich final states, that may include mono-lepton, di-leptons, tri-leptons 
and four-leptons along with $\mET$ in the final states. Note that in the absence of any mixing between the SM leptons and VLL's, the all hadronic multi-jet +$\mET$ is the dominant signal. However this signal would be swamped by  
huge SM backgrounds, which leads us to consider multi-lepton final states starting with at least one charged lepton ($e/\mu$) as a more useful signal for this model.  
We shall perform the analysis for the collider signals based on five benchmark points (BP1, BP2, BP3, BP4, BP5) given 
in Table \ref{table:masses}. 
We tabulate the two-body and three-body decay branching ratios of the charged and neutral VLLs in Appendix \ref{app : B} 
(Table \ref{BR:charged} and Table \ref{BR:neutral1} and \ref{BR:neutral2}).  We must however note that for all 
benchmark points, the relative mass splittings among the mother and daughter particles of the VLL in the cascade 
decays are not very large, leading to a somewhat compressed spectrum. This would imply relatively softer 
decay products in the final state for some of the benchmark points leading to challenges in signal-background 
discrimination, as we will see in our analysis. We therefore try to use machine learning methods in a few channels to 
check what kind of improvement one may achieve over the traditional cut-based analysis.

To check that our choice of benchmark points do not conflict any existing LHC analysis in a given leptonic channel, we 
validate these points against existing multi-lepton searches by the ATLAS and CMS collaborations. For example,
the final state containing $1\ell + \mET$ originated from the decay of $W'$ has been 
studied by ATLAS at $\sqrt{s} = 8$ TeV and integrated luminosity of 20.3 fb$^{-1}$ in \cite{ATLAS:2014wra},  
with a similar study carried out by CMS both at $\sqrt{s} = 8$ TeV and 
13 TeV \cite{Khachatryan:2014tva,Sirunyan:2018mpc}. The 
electroweak production of charginos and sleptons decaying into final states with $\ell^+\ell^- + \mET$ 
has been 
explored by ATLAS at 13 TeV LHC \cite{Aad:2019vnb}. Similarly search results for $3\ell + \mET$ final state arising 
from the decay of pair produced  chargino-neutralino with degenerate masses (with mass splitting at electroweak scale) 
has been reported by ATLAS at $\sqrt{s} = 13$ TeV in Ref.~\cite{ATLAS:2020ckz} (\cite{Aad:2019vvi}). In addition, 
search for the more robust final state with four or more charged leptons in supersymmetric framework by ATLAS at 
13 TeV LHC has also been summarised in Ref.~\cite{Aaboud:2018zeb}. Finally, a detailed study of the multi-lepton 
final state coming from the decay of doubly- and singly-charged Higgs bosons has also been performed by ATLAS 
at 13 TeV LHC \cite{Aad:2021lzu}. Although the above studies are in context of other BSM scenarios, the overlap 
with our signal topology allows us to use these studies to check whether our representative points are allowed or not. 
The checks have been performed for the five benchmarks using the mono-lepton \cite{Park:2020yyn}, 
di-lepton \cite{Araz:2020dlf} and multi-lepton \cite{Sirunyan:2017lae} searches in 
{\tt Madanalysis5} \cite{Araz:2020lnp,Araz:2019otb,Conte:2018vmg,Dumont:2014tja,Conte:2014zja,Conte:2012fm}.


	
For the chosen benchmark points, we implement the model using FeynRules~\cite{Alloul:2013bka},
which gives the required UFO that is fed in {\tt MG5aMC@NLO}~\cite{Alwall:2014hca} to generate the signal 
and background events with the cross-section at the leading order (LO). 
The LO production cross-sections at the LHC for signal and SM backgrounds are calculated using
the {\tt NNPDF3.0} parton distribution functions (PDF). To simulate the showering and hadronisation, the parton level 
events are passed through {\tt Pythia8}~\cite{Sjostrand:2014zea}. Finally, we implement the detector effects in our analysis 
using the default CMS detector simulation card for LHC available in {\tt Delphes-3.4.1} \cite{deFavereau:2013fsa}. 
For jet reconstruction, the {\texttt anti-$k_t$} clustering algorithm has been used throughout. Besides the traditional 
cut-based analysis to compute the signal significance, more sophisticated technique, {\em i.e.} Decorrelated Boosted 
Decision Tree (BDTD) algorithm is used for improvement. For such analysis, the Toolkit for Multivariate Data 
Analysis (\texttt{TMVA}) package~\cite{Hocker:2007ht} has been used. Details of the package will be discussed later 
in subsection \ref{monolep}. The signal significance $\mathcal{S}$ is derived using 
$\mathcal{S} = \sqrt{2\Big[(S + B) \log\Big(\frac{S + B}{B}\Big)- S\Big]}$, with $S (B)$ denoting the number of 
signal (background) events surviving the cuts applied on the kinematical variables. The number of signal ($S$) and the 
background ($B$) events can be calculated as:
\bea
S(B) = \sigma_{S(B)} \times \mathcal{L} \times \epsilon_{S(B)} \,,
\eea
where $\sigma_{S(B)},~\mathcal{L},~\epsilon_S (\epsilon_B)$\footnote{$\epsilon = \frac{ \rm Number ~of ~events
~surviving ~after~ applying ~cuts}{\rm Total~ number~ of ~events~ generated }.$} denote the signal (background) 
cross-section, integrated luminosity and signal (background) cut-efficiency respectively. Following this strategy, 
let us proceed to perform the collider analysis of the aforementioned channels at 14 TeV high luminosity (HL)-LHC.

\subsection{Mono-lepton final state}~\label{monolep}

To include all possible  processes leading to a signal containing mono-lepton and missing transverse 
energy ($\mET$) in the final state, we take into account the pair production and associated production of the VLL's:
\begin{eqnarray} 
&& p p \to E_i^+ E_j^- \nonumber \\
&& p p \to N_k N_m  \nonumber \\
&& p p \to E_i^\pm N_k
 \label{inclusive}
\end{eqnarray}
where $i,j = 1...4$ and $ k,m = 1...8$. 
Following the decay cascades listed in Appendix \ref{app : B}, of all the final states arising from the decay of VLLs, 
we choose the final states yielding one lepton $\ell_1$ (electron or muon) with a minimum transverse 
momentum $p_T^{\ell_1} > 10$ GeV and reject any additional lepton with $p_T^{\ell} \ge 10$ GeV. We also put veto on 
any hadronic activity by rejecting all jets with $p_T^{j} > 20$ GeV. This ensures that the signal consists of a single
charged lepton with $p_T^{\ell_1} > 10$ GeV and $\mET$. 

The irreducible SM background for this final state is the $W$ boson mediated $p p \to \ell \nu$, with $\ell = e, \mu$. 
There can be additional contributions from the di-boson productions, such as $W^\pm \,Z, \, W^+ \,W^- $ and $ZZ$, 
which yield one or more charged leptons in the final state if only leptonic decays are allowed and where the additional 
charged leptons are missed. Similarly, the multi-jet QCD background could 
also be a source of the mono-lepton background provided one of the jets is mis-tagged as a charged lepton. Although 
the probability of mis-identifying a jet as a charged lepton is rather small ($\lesssim 10^{-5}$) \cite{Aad:2009wy}, 
the sheer size of the QCD cross section makes it non-negligible. However, we also require a large missing 
transverse energy in the final state along with a jet veto which helps in suppressing the QCD background to 
negligible values. Thus in the study, we can afford to ignore this background completely. 

To generate the signal and backgrounds, we apply the following criterion to identify the isolated objects 
($\Delta \, R_{ij}>0.4$):
\begin{eqnarray}
p_{T}^j > 20 ~{\rm GeV},~~~~|\eta_j| < 5.0, \nonumber \\
p_{T}^{\ell} > 10 ~{\rm GeV},~~~~|\eta_\ell| < 2.5,
\label{basic_cuts}
\end{eqnarray}
In Table~\ref{tab:bp_lhc:monolep}, the cross-sections of the signal ($p p \rightarrow  1 \ell + \mET$) for the chosen 
benchmark points 
are tabulated for 14 TeV LHC.

\begin{table}[htpb!]
\begin{center}\scalebox{0.99}{
\begin{tabular}{|c|c|}
\hline
Benchmark points  & 
cross-section (fb) \\ \hline \hline
 BP1  &  71.57 \\
 BP2  &  53.30 \\
 BP3  &  22.25 \\
 BP4  &   4.74 \\
 BP5  &   2.58 \\ \hline \hline
\end{tabular}}
\end{center}
\caption{The LO cross-sections
for the signal $p p \rightarrow 1 \ell + \mET$ . }
\label{tab:bp_lhc:monolep}
\end{table}
\begin{figure}[htpb!]{\centering
\subfigure[]{
\includegraphics[height = 6 cm, width = 8 cm]{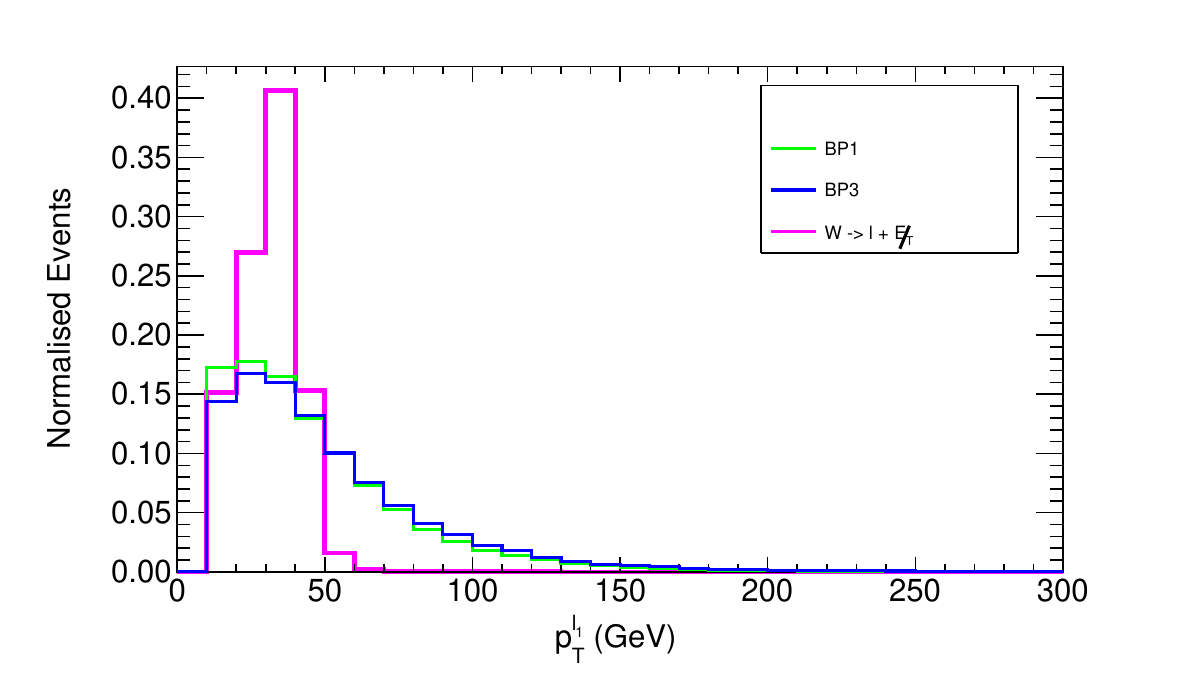}}
\subfigure[]{
\includegraphics[height = 6 cm, width = 8 cm]{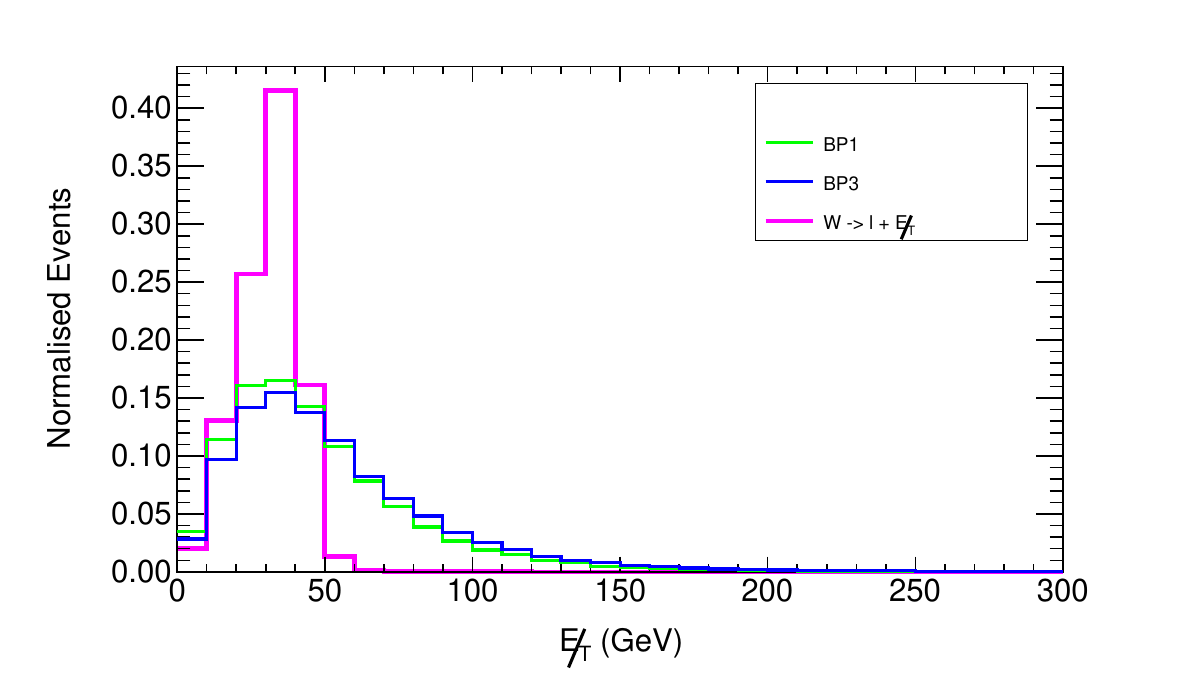}} \\
\subfigure[]{
\includegraphics[height = 6 cm, width = 8 cm]{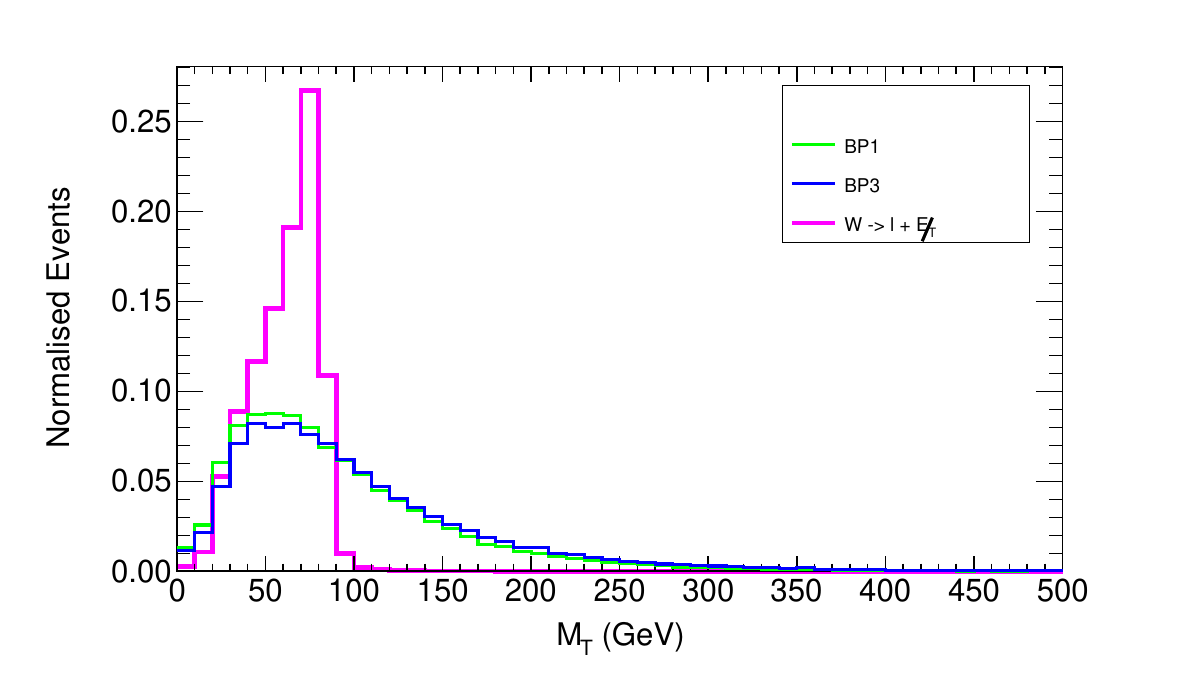}}
\subfigure[]{
\includegraphics[height = 6 cm, width = 8 cm]{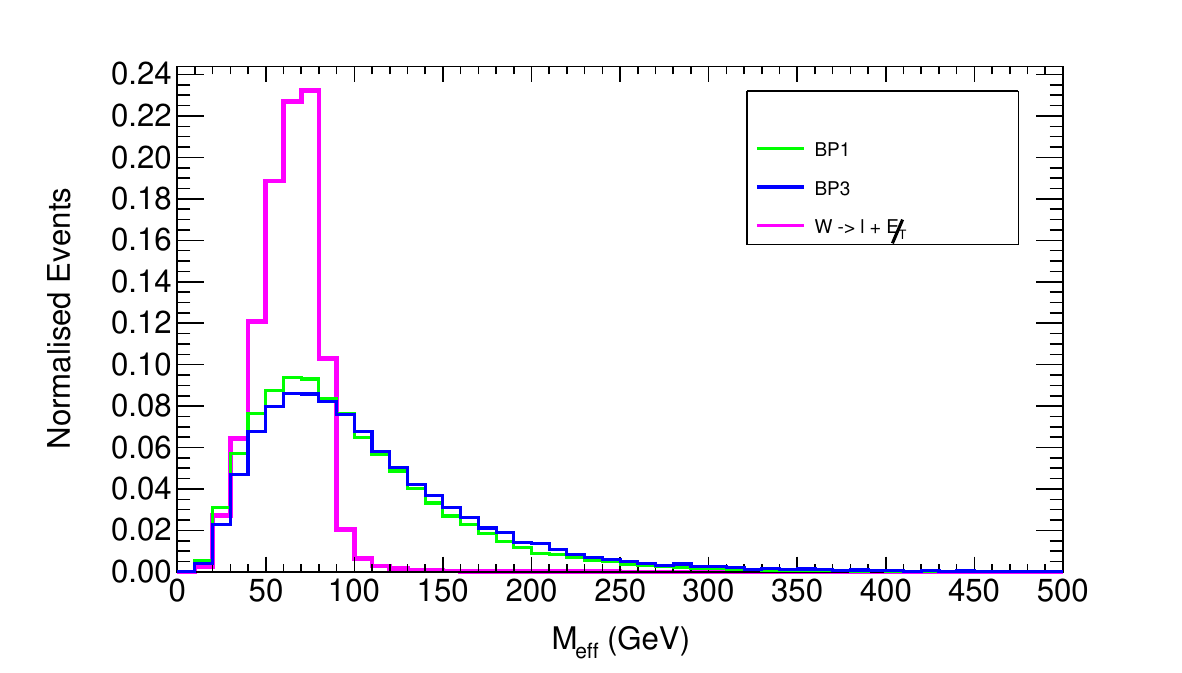}}} 
\caption{ Normalised distributions of (a) $p_T^{\ell_1}$, (b) $\mET$, (c) $M_T$, and (d) $M_{\rm eff}$ for 
$1 \ell + \mET$ channel at 14 TeV HL-LHC. }
\label{distribution-monolep}
\end{figure}
To perform the cut-based analysis, we apply the following selection cuts on chosen kinematic variables to 
disentangle the signal from SM backgrounds:{\footnote{We plot the relevant kinematic distributions for 
only BP1 and BP3 as representative points while only the dominant SM background via $p p \to W^\pm \to \ell^\pm \mET$
which is $\sim\mathcal{O}(10^4)$ bigger than the rest is shown.}}
\begin{itemize}
\item $A_1$ : From Fig.\ref{distribution-monolep}(a) it can be seen that the $p_T$-distribution of the lepton for 
the SM background coming from the 
decay of $W$ boson has the sharp Jacobian peak at $\sim M_W/2$, whereas the corresponding distribution is 
smeared for the signal where the charged lepton comes from the cascade decays of the heavy VLLs. However, a large 
part of the signal overlaps where the background peaks. 
Thus we demand that the charged lepton has a minimum transverse momentum  $p_T^{\ell} > 20$ GeV to 
exclude a significant part of the sharply peaked background (magenta line) without losing too many signal events. 

\item $A_2$ : A lower cut on the missing transverse energy, {\em i.e.} $\mET > 60$ GeV helps to reduce the 
background drastically as the $\mET$  distribution for the background (magenta line) in Fig.\ref{distribution-monolep}(b) peaks 
sharply at lower value than the signal.  Unlike the background where the neutrinos carry most of the imbalanced missing 
transverse energy, the signal gets contributions from neutrinos as well as from the much massive DM 
candidate ($N_1$) which is the end-product of all cascades giving rises to a much larger $\mET$ in the signal distribution.
\item $A_3$ : The next kinematic variables used for separating signal from background is {\em transverse mass} 
$(M_T)$ which is defined as \cite{Sirunyan:2018mpc},
\begin{equation}
 M_T = \sqrt{2 p^{\ell_1}_T \mET ~(1- \cos~ \Delta \phi_{\ell_1,\mET}~)}
\end{equation} 
where $\Delta \phi_{\ell_1, \mET}$ is the azimuthal angle between the lepton and $\mET$. 
 In Fig.\ref{distribution-monolep}(c), the $M_T$ distribution sharply peaks around $M_W$ for the background 
 as expected, while the signal has a comparatively smeared distribution. Thus we demand 
$M_T > 100$ GeV to eliminate the sharp background peak which in turn enhances the signal significance. 

\item $A_4$ :  Distribution of $M_{\rm eff}$ is depicted in Fig.\ref{distribution-monolep}(d). $M_{\rm eff}$ is defined 
by the scalar sum of the lepton $p_T$ and $\mET$.  We find that putting a lower cut $M_{\rm eff} > 110$ GeV for all the 
benchmark points helps enhance the signal over background.

\end{itemize}

\begin{table}[ht!]
	\centering
		\resizebox{14cm}{!}{
	\begin{tabular}{|p{3.0cm}|c|c|c|c|p{3.0cm}|}
		\cline{2-5}
		\multicolumn{1}{c|}{}& \multicolumn{4}{|c|}{Number of Events after cuts ($\mathcal{L}=3$ ab$^{-1}$)} &\multicolumn{1}{c}{} \\ \cline{1-5}
		SM-background  
		 & $A_1$  &  $ A_2 $    &  $A_3$  & $A_4$   & \multicolumn{1}{c}{}
		\\ \cline{1-5} 
                  $W^\pm \to \ell^\pm + \mET$  &  $3.17\times10^{10}$  & $9.46\times10^{7}$ & 
                  $6.89\times10^{7}$ & $4.47\times10^{7}$   \\ \cline{1-5} 
		$W^{+}W^{-}$ &  $4.3\times10^{6}$    & 654237   &  559027 & 431371 \\ 
		           \cline{1-5}
		 $ZZ$    & 377693 & 92344 &   84679 &  70205 \\
		             \cline{1-5}          
		$W^{\pm}Z$     & 546222   & 110314 & 105889 & 86382 \\ \hline \hline
			\multicolumn{1}{|c}{Signal }  &\multicolumn{4}{|c|}{} & \multicolumn{1}{c|}{Significance ($\mathcal{S}$)}  	
			 \\ \cline{1-6}
		\multicolumn{1}{|c|}{BP1}  & 177662  & 58170 & 56161 & 47184  & \multicolumn{1}{c|}{7.0} \\ \hline
		\multicolumn{1}{|c|}{BP2}  &  133266  & 44078 & 42216 & 35382   & \multicolumn{1}{c|}{5.3} \\ \hline
		\multicolumn{1}{|c|}{BP3}   & 54678   &  20253 & 19481  & 16587   & \multicolumn{1}{c|}{2.4} \\ \hline
		\multicolumn{1}{|c|}{BP4} &  11840  & 3804 & 3568 & 2926 & \multicolumn{1}{c|}{0.4} \\ \hline 
		\multicolumn{1}{|c|}{BP5} & 6593  & 2506 & 2399  & 2051 & \multicolumn{1}{c|}{0.3} \\ \hline 		
	\end{tabular}}

	\caption{ The cut-flow for signal and backgrounds along with the significances for BP1, BP2, BP3, BP4 and BP5 at 14 TeV HL-LHC for 3 ab$^{-1}$ integrated luminosity for the $ p p \rightarrow 1 \ell +\mET$ channel. }
	\label{tab:monolep_cut}
\end{table}	
We summarize the cut flow for both signal and background and calculate the signal significances for all five 
benchmarks in Table~\ref{tab:monolep_cut}. The table also shows the efficacy of the applied cuts for enhancing the signal 
significance.
With 3 ab$^{-1}$ integrated luminosity BP1, BP2 and BP3 can be probed with significances $7.0,~ 5.3,~ 2.4$ respectively. 
Remaining two benchmarks BP4 and BP5 yield negligible significances owing to small signal cross-sections.

Having completed the cut-based analysis, we now proceed to perform the multivariate analysis (MVA) 
using {\em Decorrelated Boosted Decision Tree} (BDTD) algorithm within the Toolkit for Multivariate Data Analysis
(TMVA) framework, with the hope of improvement in signal significance compared to the cut-based one. Before 
doing the BDTD analysis of the channels, let us present a brief overview of the method.  

To classify the signal-like or background-like events, {\em decision trees} are used as classifier. One discriminating variable 
with an optimised cut value applied on it is associated with each node of the decision tree, to provide best possible separation 
between the signal-like and background-like events depending on the purity of the sample. Within TMVA, this can be done by 
tuning the BDTD variable {\texttt{NCuts}}. The training of the decision trees starts from the root node (zeroth node) and 
continues till a particular depth specified by the user is reached. This particular depth is termed as {\texttt{MaxDepth}}. 
Finally from the final nodes or the {\em leaf nodes} an event can be specified as signal or background according to 
their {\em purity}. {\footnote{The purity $p$ can be defined as : $ p = \frac{S}{S+B}$. }} An event can be tagged as 
signal (background) when $p > 0.5$ ($p < 0.5$).

Now the decision trees are termed as weak classifiers as they are  prone to statistical fluctuations of the training 
sample. To circumvent this problem, one can combine a set of weak classifiers into a stronger one and create new decision 
trees by modifying the weight of the events. This procedure is termed as {\em Boosting}. For this analysis, we 
choose {\em Adaptive boost} with the input variables transforming in a decorrelated manner, since this is very useful for 
weak classifiers. In TMVA it is implemented as {\texttt Decorrelated AdaBoost}. Several BDTD parameters like the 
number of decision trees {\texttt{NTrees}}, the maximum depth of the decision tree allowed
 {\texttt{MaxDepth}}, the minimum percentage of training events in each leaf node {\texttt{MinNodeSize}} and {\texttt{NCuts}} 
 for five benchmarks of our analysis have been tabulated in Table \ref{BDT-param}. 
 To avoid over training of the signal and background 
 samples, the results of the Kolmogorov-Smirnov test, {\em i.e.} Kolmogorov-Smirnov score (KS-score) should 
always be > 0.1\footnote{KS-score > 0.01 will also serve the purpose if it remains stable even after changing the 
internal parameters of the algorithm.} and stable.
\begin{table}[htpb!]
\begin{center}
\resizebox{16cm}{!}{
\begin{tabular}{|c|c|c|c|c|c|}
\hline
 &  \hspace{5mm} {\texttt{NTrees}} \hspace{5mm} & \hspace{5mm} {\texttt{MinNodeSize}} \hspace{5mm} & \hspace{5mm} {\texttt{MaxDepth}}~~ \hspace{5mm} & \hspace{5mm} {\texttt{nCuts}} ~~\hspace{5mm} & \hspace{5mm} {\texttt{KS-score for}}~~\hspace{5mm}\\
 & & & & & {\texttt{Signal(Background)}} \\
\hline
\hline
\hspace{5mm} BP1 \hspace{5mm} & 120 & 4 \% & 3.0 & 40 & 0.328(0.256) \\ \hline
\hspace{5mm} BP2 \hspace{5mm} & 100 & 1 \% & 3.0 & 40 & 0.017(0.502) \\ \hline
\hspace{5mm} BP3 \hspace{5mm} & 120 & 4 \% & 3.0 & 40 & 0.176(0.105)  \\ \hline
\hspace{5mm} BP4 \hspace{5mm} & 120 & 4 \% & 3.0 & 40 & 0.102(0.607) \\ \hline
\hspace{5mm} BP5 \hspace{5mm} & 120 & 4 \% & 3.0 & 40 & 0.213(0.087) \\ \hline
\end{tabular}}
\end{center}
\caption{Tuned BDT parameters for BP1, BP2, BP3, BP4 and BP5 for the $1 \ell + \mET$ channel.}
\label{BDT-param}
\end{table}
 
\begin{figure}[tp!]{\centering
\subfigure[]{
\includegraphics[width=2.8in,height=2.55in]{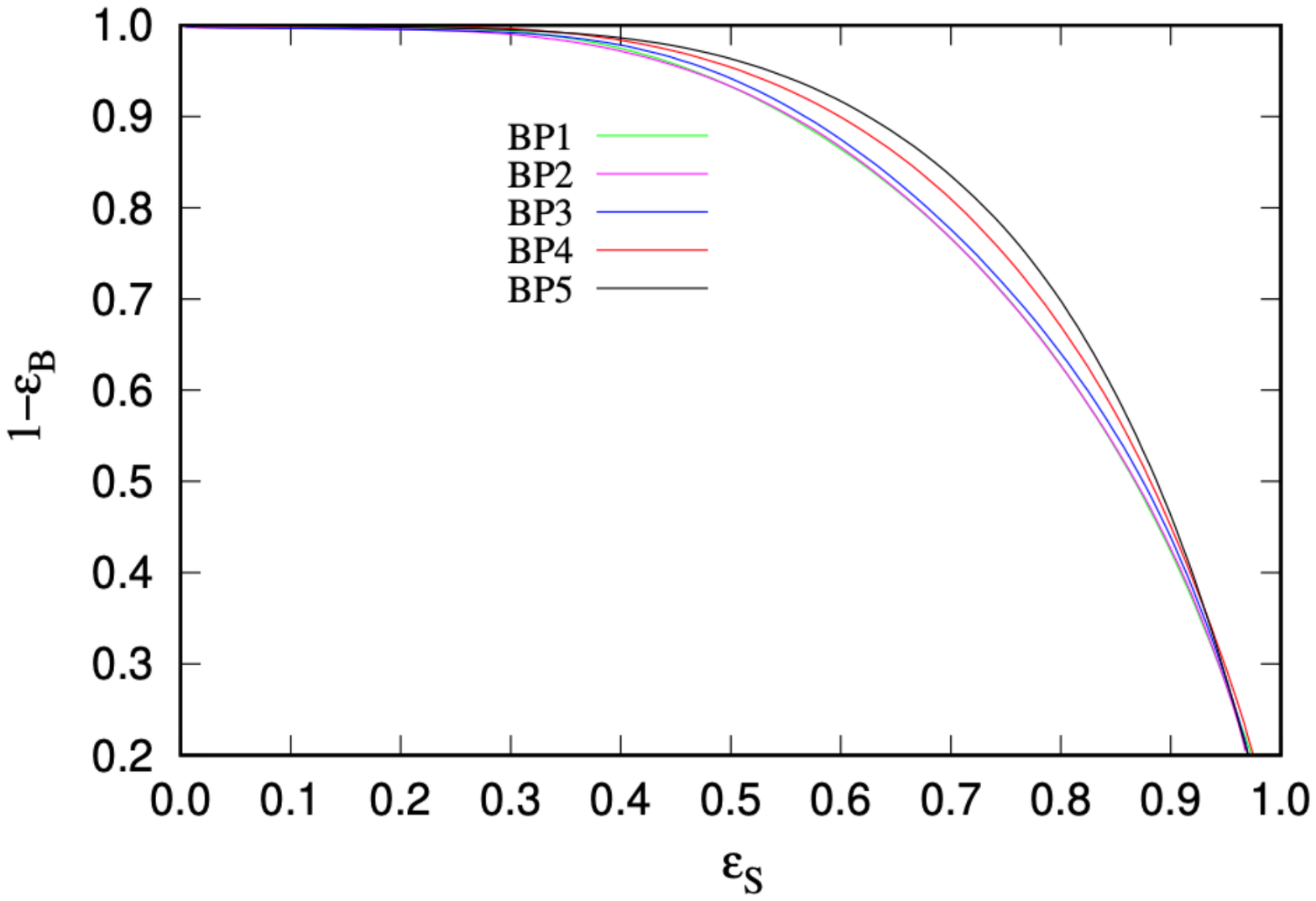}}
\subfigure[]{
\includegraphics[width=2.8in,height=2.45in]{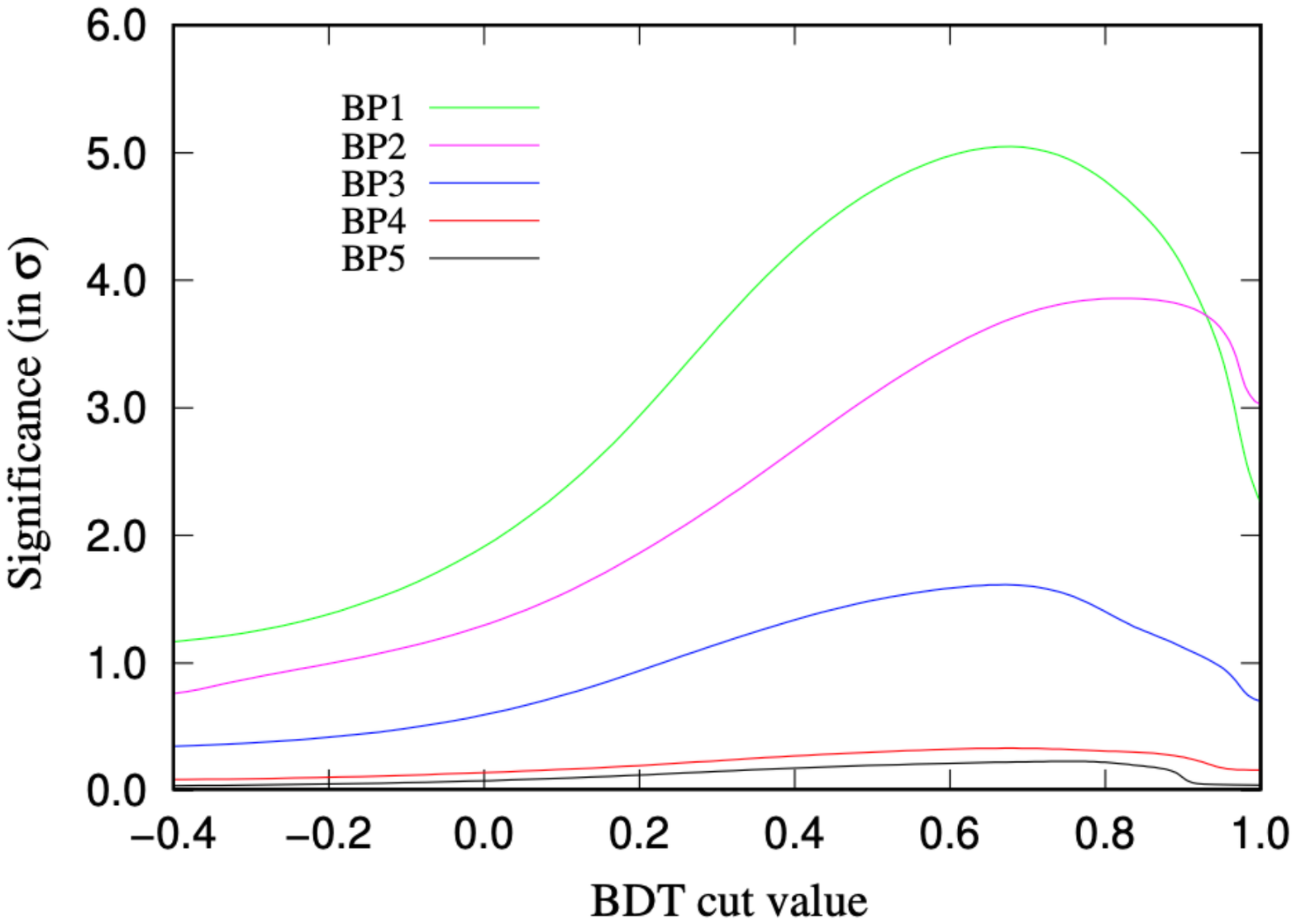}}}
\caption{(a) ROC curves for chosen benchmark points for $1 \ell + \mET$ channel. 
(b) Variation of significance with BDT-score for $1 \ell + \mET$ channel.}
\label{ROC-BDTScore-monolep}
\end{figure}

According to the degree of discriminatory power between the signal and backgrounds, following are the
kinematic variables of importance :
\bea
 M_T,~ M_{\rm eff}, ~ \mET
\eea
 These relevant kinematic variables are constructed for each and every channel to discriminate between the signal 
 and the backgrounds.  
\begin{figure}[htpb!]{\centering
\subfigure[]{
\includegraphics[height = 6 cm, width = 8 cm]{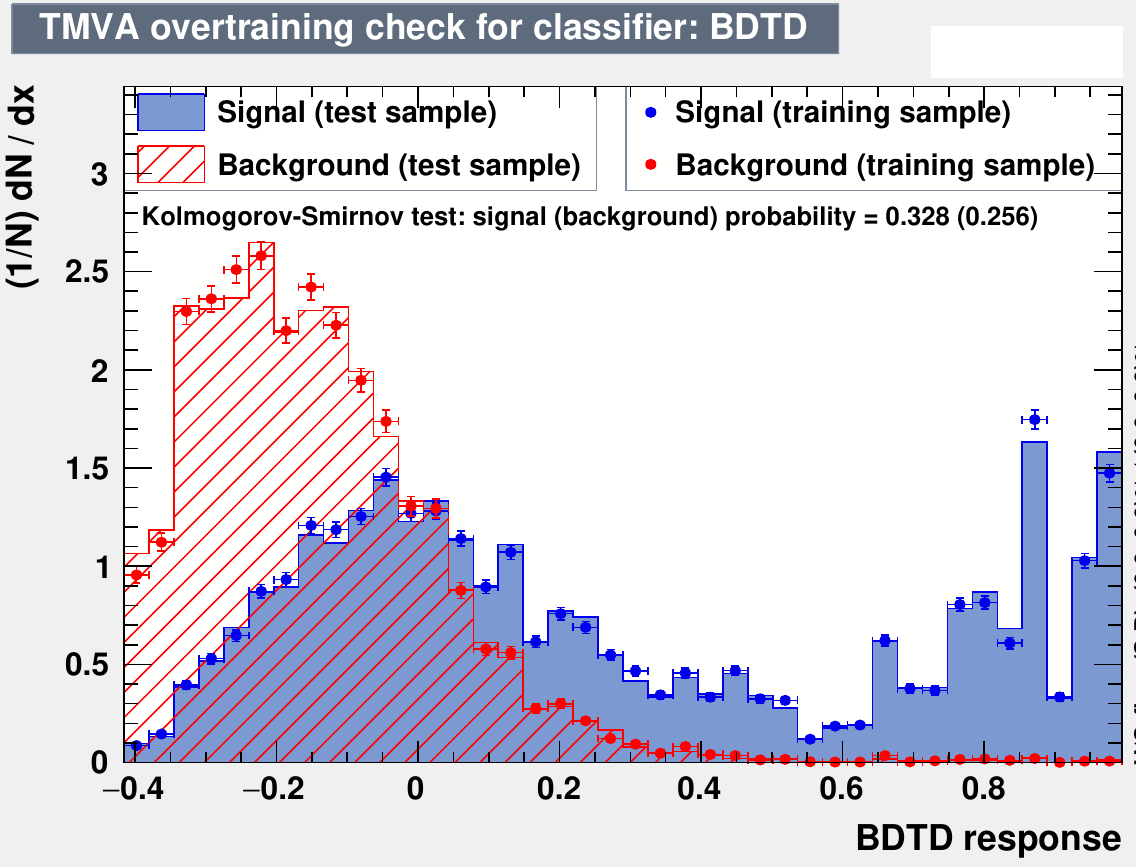}}
\subfigure[]{
\includegraphics[height = 6 cm, width = 8 cm]{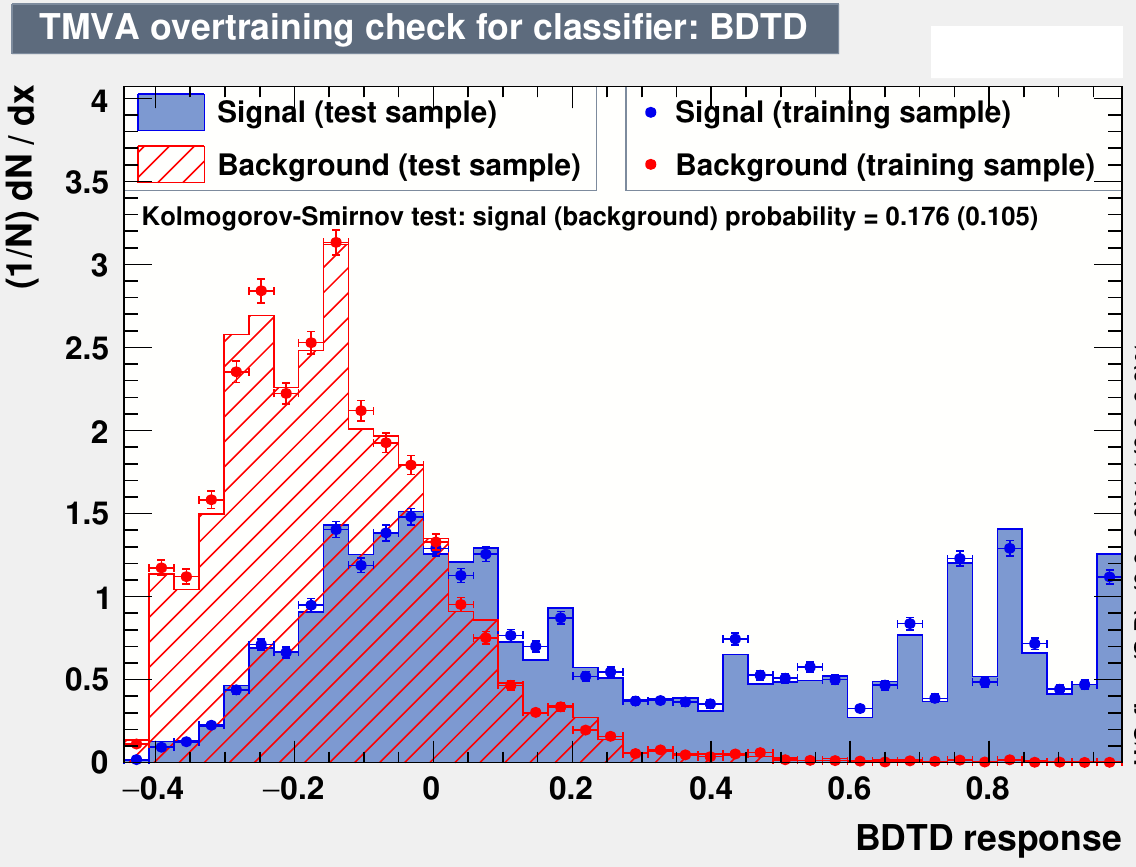}} 
           }
\caption{ KS-scores corresponding to (a) BP1 and (b) BP3 for $1 \ell + \mET$ channel.}
\label{KSScore-monolep}
\end{figure}
Fig.\ref{KSScore-monolep} shows the KS-scores for both signal (blue distribution) and backgrounds (red distribution) 
for two representative benchmark points BP1 and BP3. For convenience, we have tabulated the KS-scores for 
both signal and background in the sixth column of Table \ref{BDT-param} for all benchmark points. To make the KS-score stable, one can tune the 
BDTD parameters given in Table \ref{BDT-param}. With the aforementioned discriminating variables at hand, we tune 
the {\em BDT cut value (BDT score)} in such a way that the significance is maximized. 
We plot the Receiver's Operative Characteristic (ROC) curve for all benchmarks in 
Fig.\ref{ROC-BDTScore-monolep}(a)\footnote{ROC curve is a plot of signal efficiency ($\epsilon_S$) vs. efficiency of 
rejecting the backgrounds $(1-\epsilon_B)$, $\epsilon_B$ being the background efficiency.},  which classifies the 
degree of rejecting the backgrounds with respect to the signal. Variation of the signal significance with BDT cut value for all the 
benchmarks are shown in Fig.\ref{ROC-BDTScore-monolep}(b). It can be clearly seen that the signal significance attains a 
maximum value for each benchmark at a particular value of BDT score.

Signal and background yields with 3 ab$^{-1}$ integrated luminosity after performing BDTD analysis have been 
tabulated in Table \ref{1lmet:lhc}. The significances for BP1, BP2, BP3, BP4 and BP5 are 7.8, 5.9, 2.6, 0.5 and 0.4 respectively. It becomes quite clear that compared to the cut-based analysis, the signal 
significances get typical improvements of $11.4\, \%,\, 11.3 \,\%,\, 8.3\,\%,\, 25\,\%$ and $33.3\, \%$ for BP1, BP2, BP3, 
BP4 and BP5 respectively.  An integrated luminosity of $\sim$1232 fb$^{-1}$ is required to achieve 5$\sigma$ significance for BP1 after performing the BDTD analysis.

\begin{center}
\begin{table}[htb!]
\begin{tabular}{|c|c|c|c|}\hline
 Benchmark Point & Signal Yield & Background Yield & Significance \\ \hline 
BP1 & 71148 & $8.24\times10^7$  & 7.8 \\ \hline
BP2 & 47827 & $6.62\times10^7$ & 5.9 \\ \hline
BP3 & 20514 & $6.25\times10^7$ & 2.6 \\ \hline
BP4 & 3816   & $6.75\times10^7$ & 0.5 \\ \hline
BP5 & 1995 &  $2.52\times10^7$ & 0.4 \\ \hline
 
\end{tabular}
\caption{ The signal and background yields at 14 TeV-LHC with 3 ab$^{-1}$ integrated luminosity for BP1,BP2, 
BP3, BP4 and BP5 along with signal significances for the $ p p \rightarrow 1 \ell +\mET$ channel after performing 
the BDTD analysis. }
\label{1lmet:lhc}
\end{table}
\end{center}


\subsection{Di-lepton final state}
\label{dilep}

We now consider the final states containing same or different flavour and opposite sign (OS) di-leptons along with 
$\mET$ that can arise from the following subprocesses in our model:
\begin{eqnarray}\nonumber
 p p \to E^{+}_i E^{-}_j, E^{\pm}_{i,j} \to \ell^{\pm} N_1 \\
 p p \to N_1 N_k, N_k \to N_1 \ell^{+} \ell^{-}
\end{eqnarray}
where $i,j = 1,2,...4, k = 2,3,...8$. The dominant signal contribution comes from the pair production of the 
charged VLLs followed by their decay to DM and a lepton. Production of the 
vector like neutrino along with the DM can also give rise to the similar final state albeit small cross-section. However 
for the sake of completeness we take into account all such processes that may give rise to a di-lepton final state.
The major SM background for the signal comes from the inclusive $2\ell  + \mET$ process which includes contributions from $W^{+}W^{-}$ and $ZZ$ pair production. 
 Due to large cross-section, $t\bar{t}$ followed by the leptonic decay of top-quark (leading to $2 b + 2\ell + \mET$ 
 final state) also contributes as one of the major background. 
Even after a b-jet veto along with a jet-veto, the small fraction of events surviving from the $t \bar{t}$ process
can still lead to a significant number of events in the $2 \ell  + \mET $ final state. In addition, processes with smaller cross-sections such as $W^{\pm}Z$, and
$W^+W^-Z$ followed by the leptonic decay of $W^\pm$ and $Z$ can also be 
a possible source of background for the $2 \ell+\mET$ final state, if one or more leptons escape detection. For the analysis, 
we consider the above three SM subprocesses as major contributions to the SM background. In 
Table \ref{tab:bp_lhc:dilep}  the LO signal 
cross-sections for the di-lepton final state are tabulated for all our benchmark points. 


\begin{table}[htpb!]
\begin{center}\scalebox{0.99}{
\begin{tabular}{|c|c|}
\hline
Signal  & Cross-section (fb) \\ \hline \hline
 BP1  &  14.48 \\
 BP2  &  7.39 \\
 BP3  &  3.19 \\
 BP4 &   1.82 \\
 BP5 &  0.58 \\ \hline \hline
\end{tabular}}
\end{center}
\caption{The LO cross-sections 
for the signal $p p \rightarrow 2 \ell + \mET$.}
\label{tab:bp_lhc:dilep}
\end{table}


%
To generate the signal and backgrounds we apply the same set of generation cuts as mentioned in 
subsection \ref{monolep}. We select events with exactly two charged leptons with $p_T^{\ell} > 10$ GeV and 
$|\eta_{\ell}| < 2.5$ and reject any additional lepton with $p_T^{\ell} > 10$ GeV. To ensure a hadronically quiet final state, 
we veto all light-jets, $b$-jets and $\tau$-jets with $p_T > 20$ GeV. We then analyse the signal containing OS di-leptons 
and compute the signal significance using traditional cut-based method. To differentiate our signal from 
the SM background, we focus on the following kinematic variables: $p_T^{\ell_1}, p_T^{\ell_2}, \mET$ and invariant mass 
of two OS same or different flavoured leptons $M_{\ell^{+} \ell^{-}}$. We define the cuts applied on them as 
$B_1, B_2, B_3, B_4$ respectively and we describe them below :

\begin{itemize}
 
 \item $B_1$ : In Fig.~\ref{distribution-dilep}(a) and ~\ref{distribution-dilep}(b), we depict normalised $p_T$ 
 distribution for the leading and sub-leading leptons $\ell_1$ and $\ell_2$ for both signal and SM background. In can 
 be seen that
 the distributions have a significant overlap owing to their origin being from $W$ decay.  Thus we apply 
 $p_T^{\ell_1} > 20$  GeV suppress the SM backgrounds.
\begin{figure}[htp!]{\centering
\subfigure[]{
\includegraphics[height = 6 cm, width = 8 cm]{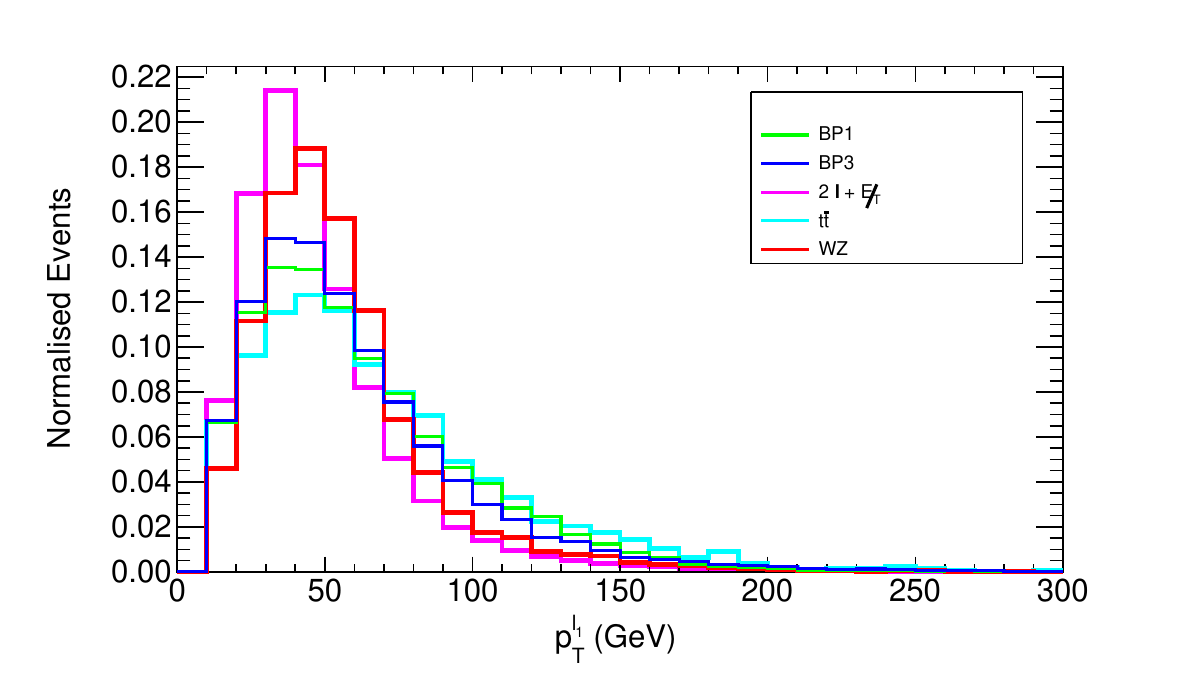}}
\subfigure[]{
\includegraphics[height = 6 cm, width = 8 cm]{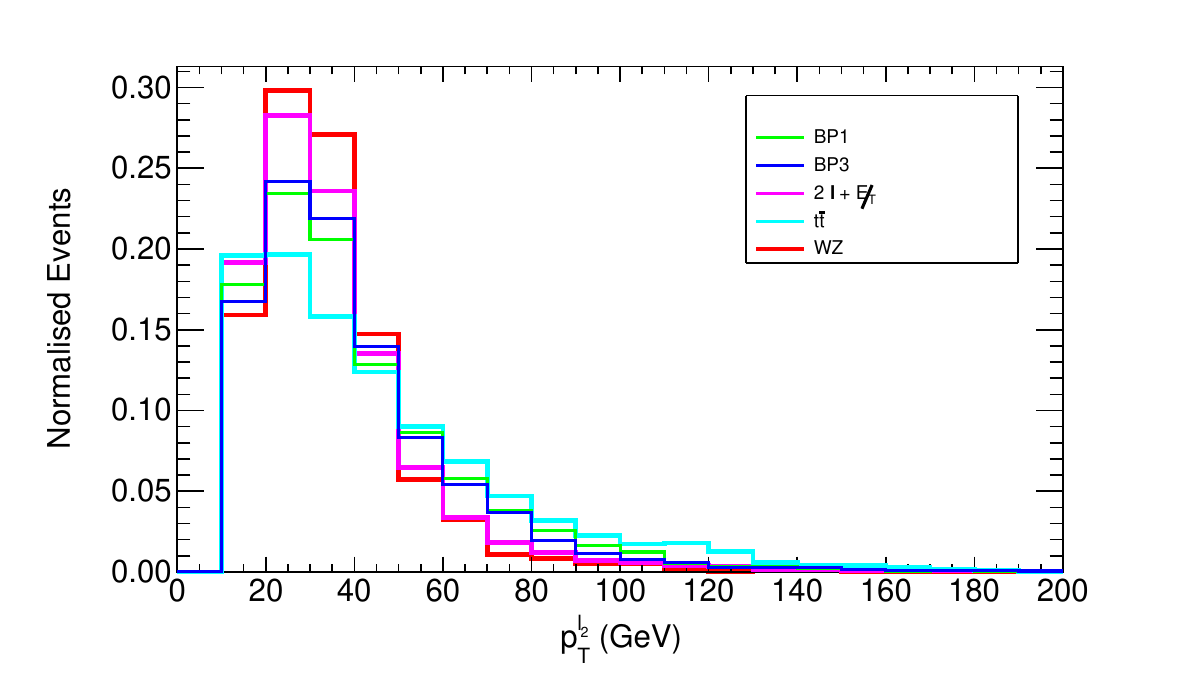}} \\
\subfigure[]{
\includegraphics[height = 6 cm, width = 8 cm]{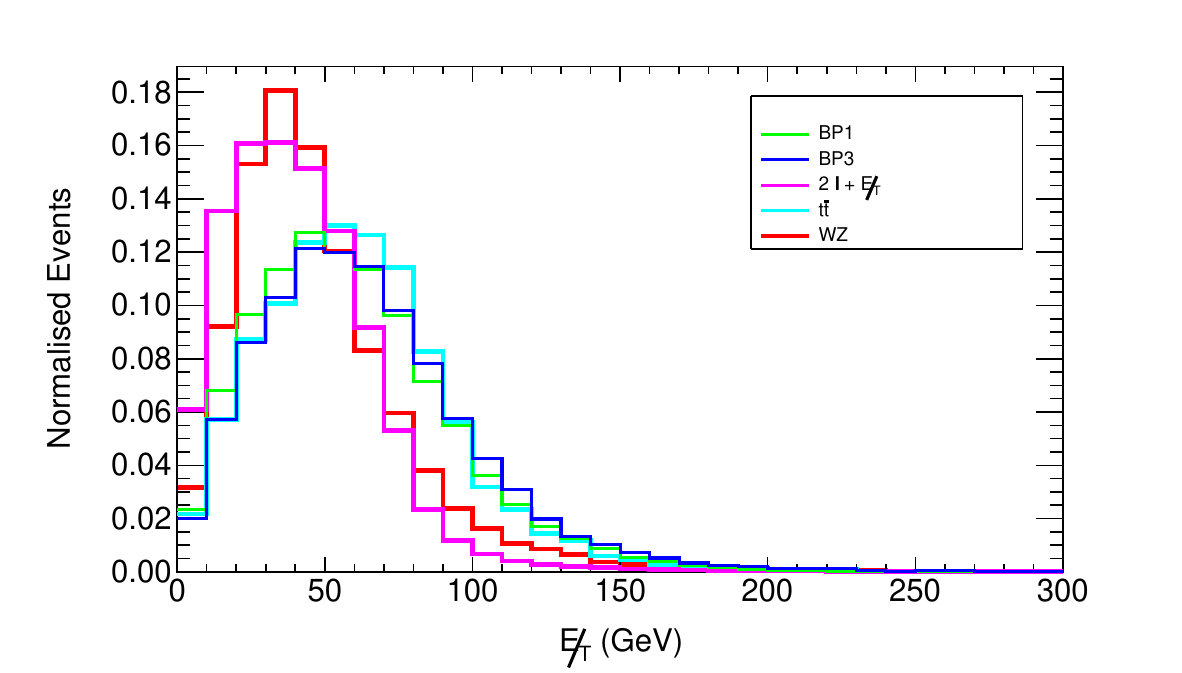}}
\subfigure[]{
\includegraphics[height = 6 cm, width = 8 cm]{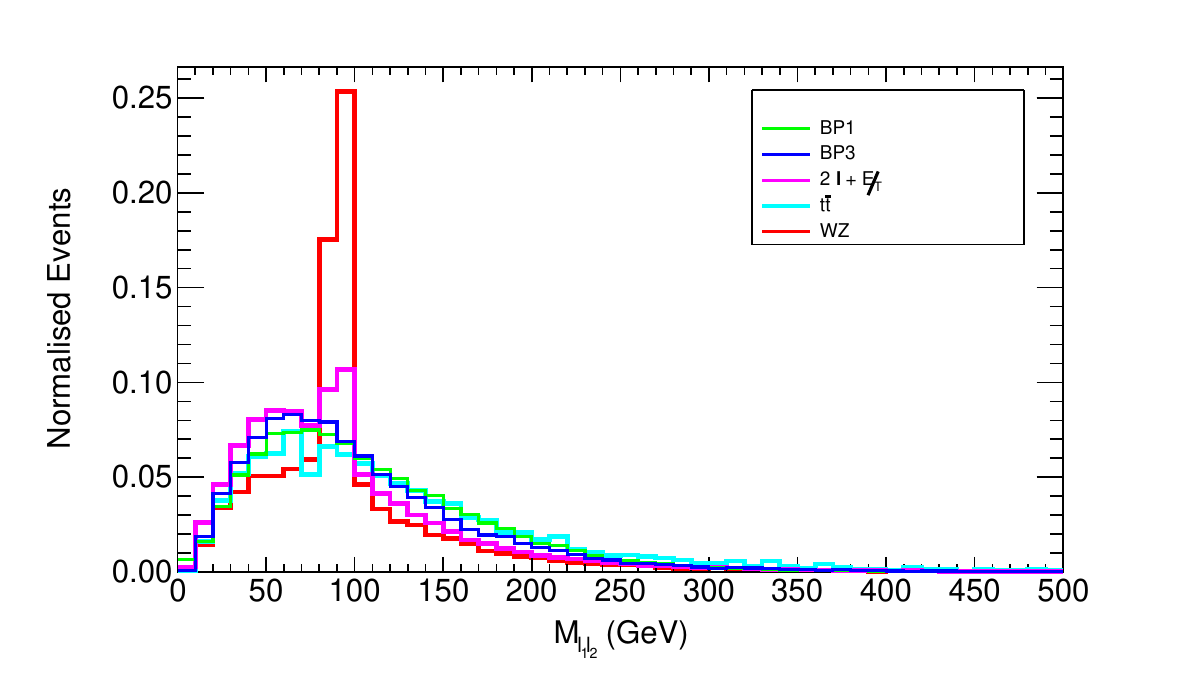}}} \\
\caption{ Normalised distributions of $p_T^{\ell_1},~p_T^{\ell_2},~\mET, ~M_{\ell_1 \ell_2}$ for $2 \ell + \mET$ 
channel at 14 TeV HL-LHC.}
\label{distribution-dilep}
\end{figure}

\item $B_2$ : The normalised distribution of missing transverse energy is shown in Fig.~\ref{distribution-dilep}(c).  The 
$\mET$ distributions for the BP1 and BP3 (green and blue lines) are much harder as in the mono-lepton case. Thus 
we demand $\mET > 40$ GeV, which helps to reduce the $2 \ell+\mET$ background. As the mass splitting 
between the VLLs become smaller for heavier DM, the $\mET$ distribution is shifted towards the softer side.
\item $B_3$ :  The normalised distribution of $M_{\ell^{+} \ell^{-}}$ is shown in Fig.\ref{distribution-dilep}(d). 
The distribution for the $WZ$ background (red line) shows a peak at $M_Z$, since two same flavour opposite sign 
leptons out of the three in the final state, originate from the $Z$-boson decay.  As the signal does not have a $Z$ peak 
in its distribution, we reject events $ 75 < (M_{\ell^+ \ell^-})_{1,2} < 105$ GeV to exclude the $Z$-peak.
This cut helps in suppressing the $WZ$ background. Along with, we also demand $(M_{\ell^+ \ell^-})_{1,2} > 12$ GeV
to reduce the Drell-Yan background contribution~\cite{Chatrchyan:2014aea}.
\end{itemize}

We sum up the number of surviving signal and background events after applying the aforementioned cuts 
in Table~\ref{tab:sig_lhc}. It can be  seen that with 3 ab$^{-1}$ integrated luminosity, the significance reach for BP1 and 
BP2 is 6.7 and 2.5 respectively. The search prospect of this channel at the same integrated luminosity for BP3, BP4 
and BP5 is considerably poor with the signal significances being 1.3, 0.8 and 0.3 respectively, owing not only to the small 
signal production cross-sections but also to the huge SM background which is not reducible. Note that for the di-lepton 
channel, the signal and background distributions have significant overlap in most kinematic variables, which makes it 
somewhat difficult to suppress the backgrounds without reducing the signal events.  
		
\begin{table}[ht!]
	\centering
		\resizebox{12cm}{!}{
	\begin{tabular}{|p{3.0cm}|c|c|c|p{3.0cm}|}
		\cline{2-4}
		\multicolumn{1}{c|}{}& \multicolumn{3}{|c|}{Number of Events after cuts ($\mathcal{L}=3$ ab$^{-1}$)} & \multicolumn{1}{c}{} \\ \cline{1-4}
		SM-background  
		 & $B_1$  &  $ B_2 $    &  $B_3$    & \multicolumn{1}{c}{}
		\\ \cline{1-4} 
                  $2 \ell + \mET$  & 2197304 & 1072431 & 751683   \\ \cline{1-4} 
		$t\bar{t}$ leptonic   & 184414 & 134828 & 105741    \\ 
		           \cline{1-4}
		$W^{\pm}Z$   & 71918 & 38373 & 17819     \\ \hline \hline
			\multicolumn{1}{|c|}{Signal }  &\multicolumn{3}{|c|}{} &\multicolumn{1}{c|}{Significance ($\mathcal{S}$)}  \\ \cline{1-5}
		\multicolumn{1}{|c|}{BP1}  & 11623 & 8169 & 6253 &  \multicolumn{1}{|c|}{6.7}  \\ \hline
		\multicolumn{1}{|c|}{BP2}  & 5058 & 3368 & 2366 & \multicolumn{1}{|c|}{2.5} \\ \hline
		\multicolumn{1}{|c|}{BP3}  &  2152 & 1598  &  1210 & \multicolumn{1}{|c|}{1.3}  \\ \hline
		\multicolumn{1}{|c|}{BP4} &   1480 & 1059 & 794 &  \multicolumn{1}{|c|}{0.8} \\ \hline 
		\multicolumn{1}{|c|}{BP5} &  464 & 327 & 244 & \multicolumn{1}{|c|}{0.3}  \\ \hline 
		
	\end{tabular}}

	\caption{ The cut-flow for signal and backgrounds for $2 \ell + \mET$ channel along with the significance for 
	BP1, BP2, BP3, BP4 and BP5 at 14 TeV LHC for 3 ab$^{-1}$ integrated luminosity. }
	\label{tab:sig_lhc}
\end{table}

\begin{table}[htpb!]
\begin{center}
\resizebox{16cm}{!}{
\begin{tabular}{|c|c|c|c|c|c|}
\hline
 &  \hspace{5mm} {\texttt{NTrees}} \hspace{5mm} & \hspace{5mm} {\texttt{MinNodeSize}} \hspace{5mm} & \hspace{5mm} {\texttt{MaxDepth}}~~ \hspace{5mm} & \hspace{5mm} {\texttt{nCuts}} ~~\hspace{5mm} & \hspace{5mm} {\texttt{KS-score for}}~~\hspace{5mm}\\
 & & & & & {\texttt{Signal(Background)}} \\
\hline
\hline
\hspace{5mm} BP1 \hspace{5mm} & 110  & 3 \% & 2.0 & 40 & 0.231(0.44) \\ \hline
\hspace{5mm} BP2 \hspace{5mm} & 110 & 4 \% & 2.0 & 40 & 0.15(0.482) \\ \hline
\hspace{5mm} BP3 \hspace{5mm} & 110 & 3 \% & 2.0 & 40 & 0.017(0.018) \\ \hline
\hspace{5mm} BP4 \hspace{5mm} & 110 & 3 \% & 2.0 & 40 & 0.016(0.22) \\ \hline
\hspace{5mm} BP5 \hspace{5mm} & 100 & 3 \% & 2.5 & 35 & 0.914(0.462)\\ \hline
\end{tabular}}
\end{center}
\caption{Tuned BDT parameters for BP1, BP2, BP3, BP4 and BP5 for the channel $2 \ell + \mET$.}
\label{BDT-param-dilep}
\end{table}

\begin{center}
\begin{table}[htb!]
\centering
\begin{tabular}{|c|c|c|c|}\hline
 Benchmark Point & Signal Yield & Background Yield & Significance \\ \hline 
BP1 & 12515 & $2.58\times10^6$ & 7.8 \\ \hline
BP2 & 6271 & $2.48\times10^6$ & 4.0 \\ \hline
BP3 & 2552 & $2.39\times10^6$ & 1.7 \\ \hline
BP4 & 1272 & $1.97\times10^6$ & 0.9 \\ \hline
BP5 & 507 & $2.53\times10^6$ & 0.32 \\ \hline
 
\end{tabular}

\caption{ The signal and background yields at 14 TeV-LHC and 3 ab$^{-1}$ integrated luminosity for BP1, BP2, 
BP3, BP4 and BP5 along with signal significances for the $ p p \rightarrow 2 \ell+\mET$ channel after the BDTD analysis. }
\label{2lmet:lhc}
\end{table}
\end{center}
\begin{figure}[htpb!]{\centering
\subfigure[]{
\includegraphics[height = 6 cm, width = 8 cm]{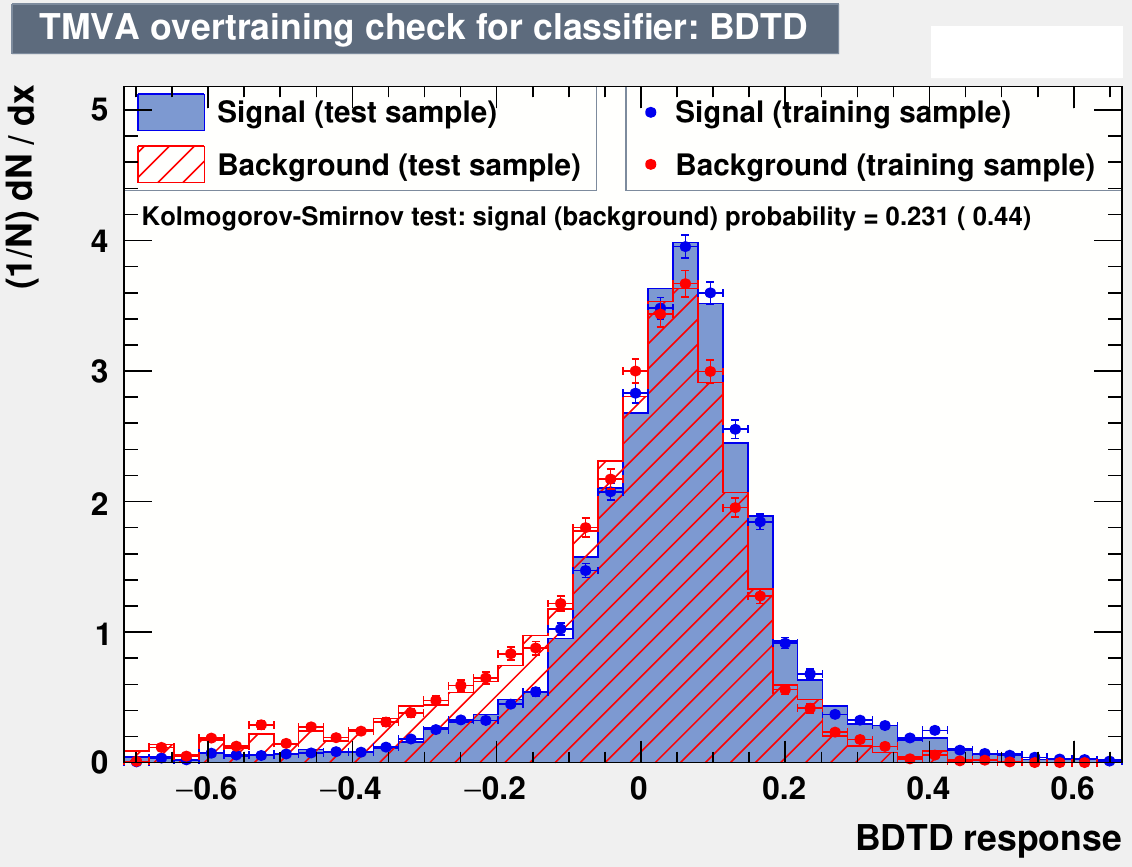}}
\subfigure[]{
\includegraphics[height = 6 cm, width = 8 cm]{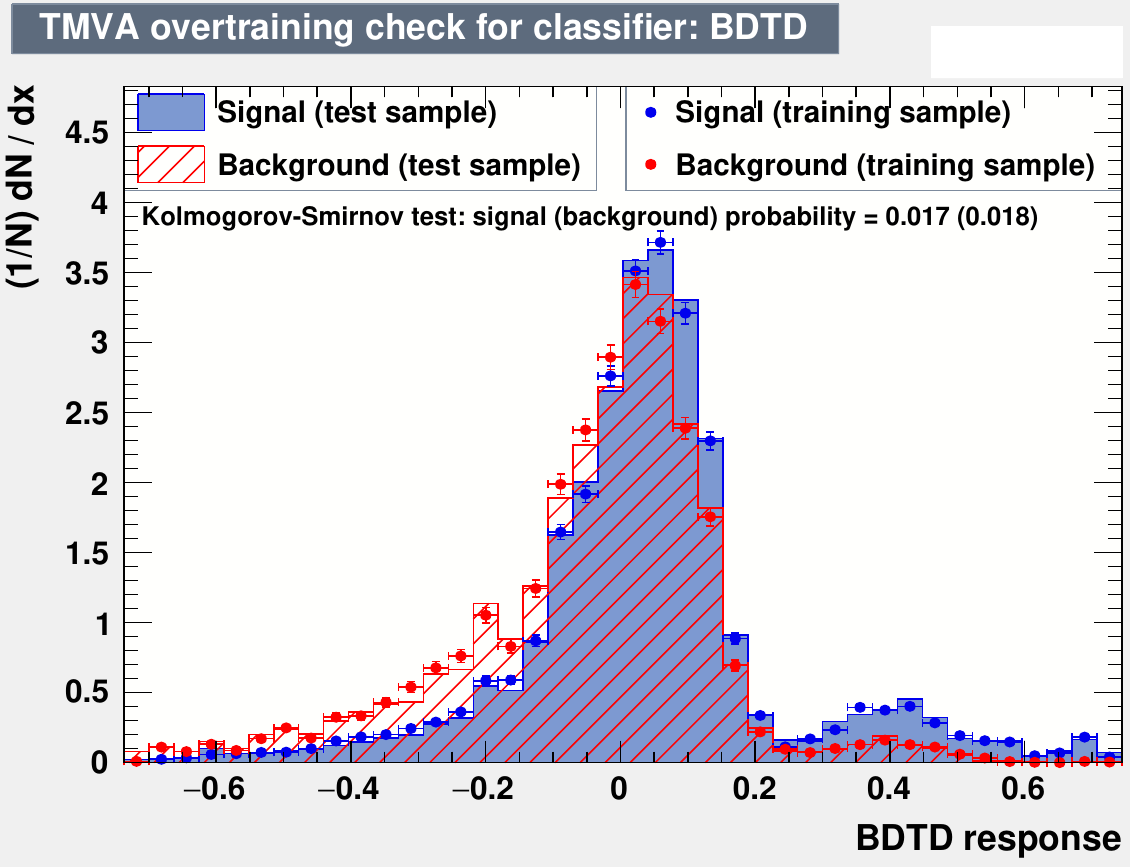}} 
                               } 
\caption{KS-scores corresponding to (a) BP1 and (b) BP3 for $2 \ell + \mET$ channel.}
\label{KSScore-dilep}
\end{figure}

\begin{figure}[htpb!]{\centering
\subfigure[]{
\includegraphics[width=3in,height=2.45in]{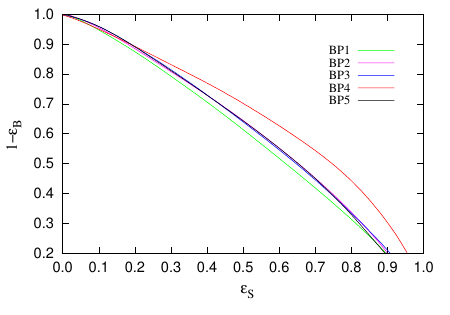}}
\subfigure[]{
\includegraphics[width=3.1in,height=2.46in]{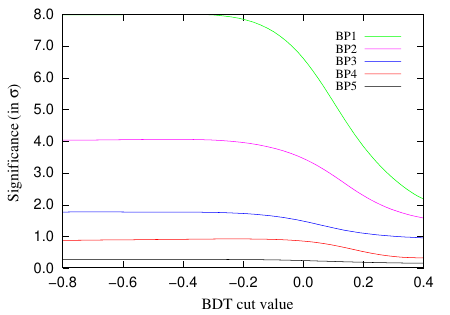}}}
\caption{ (a) ROC curves for chosen benchmark points for $2 \ell + \mET$ channel. (b) BDT-scores corresponding 
to BP1, BP2, BP3, BP4 and BP5 for $2 \ell + \mET$ channel.}
\label{ROC-BDTScore-dilep}
\end{figure}



After discussing the cut-based analysis, let us move on to the multivariate (BDTD) analysis, which improves 
the signal significance by enhancing the discriminatory power between the signal and the backgrounds.
For this analysis, we consider the following kinematic variables with maximal discerning ability :
\bea
M_{\ell^+ \ell^-}, ~ \mET, ~ p_T^{\ell_1}, ~\Delta R_{\ell_1 \ell_2} \,
\label{BDT:lhc}
\eea
 Using these variables we train the signal and backgrounds so that the signal significance is 
 maximized. %
 

We present the set of tuned BDT parameters for all the benchmarks in Table \ref{BDT-param-dilep} to make the 
KS-score stable following the criteria mentioned in Sec. \ref{monolep}. The KS-scores for BP1 and BP3 (both for signal 
and background) are given in Fig.\ref{KSScore-dilep}. In the sixth column of Table \ref{BDT-param-dilep} KS-scores for all benchmarks have been quoted. Having fixed 
the KS-score, we next proceed to tune the BDT score to yield maximum significance. 
Background rejection efficiency vs. signal efficiency have been plotted in the ROC curves in 
Fig.\ref{ROC-BDTScore-dilep}(a) using the aforementioned kinematic variables. From the ROC curves of the 
$2 \ell + \mET$ channel, it is evident that the background rejection efficiency is somewhat poor compared to the 
$1 \ell + \mET$ channel. The significances have been plotted against BDT score for all benchmarks in 
Fig.\ref{ROC-BDTScore-dilep}(b). 

Signal and background yields with 3 ab$^{-1}$ integrated luminosity for our chosen benchmark points along with 
the significances are listed in Table \ref{2lmet:lhc}. From Table \ref{2lmet:lhc} it can be inferred that the signal 
significance has improved a bit compared to the cut-based counter part. For BP1, BP2, BP3, BP4 and BP5 the 
improvements in signal significance are $16.4\%, \, 60.0\%, \, 30.8\%, \, 12.5\%$ and $6.7\%$ respectively. 

\subsection{Tri-lepton final state}
\label{trilep}

The tri-lepton final state can originate from the following subprocesses :
\bea
&& p p \to E_i^\pm N_j, \nonumber \\
&& E_i^\pm \to W^\pm N_1, ~ W^\pm \to \ell^\pm \nu_\ell \nonumber \\
 && N_j \to N_1 \ell^+ \ell^-, \nonumber \\
 && {\rm with}~~ i=1,2,3,4, ~ j = 2,3,4,...,8.
\eea
We generate the events with tri-lepton final state using the same generation-level cuts and following the method 
discussed in subsection~\ref{monolep}. Among all possible decay products of the pair produced neutral and 
charged VLLs, we select only those events which have three charged leptons and missing transverse energy in the 
final state. We consider $p p \to 3\ell + \mET$ with zero jets 
as the dominant irreducible SM background for our signal, which includes 
both on-shell and off-shell contributions from diboson and triboson production. 
In addition, the pair production of $Z$ boson where $ZZ \to 4\ell$ can also 
give rise to a similar final state if one of the leptons is missed. 
All LO cross-sections for this signal and backgrounds at 14 TeV LHC are given 
in Table \ref{tab:bp_lhc:trilep}\footnote{Before proceeding towards the analysis at 14 TeV, we first validate the 
chosen benchmarks using the existing search for chargino-neutralino pair production in final states with three leptons 
and missing transverse momentum at $\sqrt{s} = 13$ TeV performed by the ATLAS 
detector \cite{ATLAS:2020ckz}.}.
\begin{table}[htpb!]
\begin{center}\scalebox{0.95}{
\begin{tabular}{|c|c|c|}
\hline
Signal   & Process at LHC & LO cross-section (fb) \\ \hline \hline
 BP1 & & 1.63 \\
 BP2  & & 0.60 \\
 BP3  & $p p \rightarrow 3\ell + \mET$& 0.49 \\
 BP4 &  & 0.18 \\
 BP5 & & 0.064 \\ \hline \hline
\end{tabular}}
\end{center}
\caption{The cross-sections for the signal $p p \rightarrow 3 \ell + \mET$ .}
\label{tab:bp_lhc:trilep}
\end{table}

\begin{figure}[htpb!]{\centering
\subfigure[]{
\includegraphics[height = 6 cm, width = 8 cm]{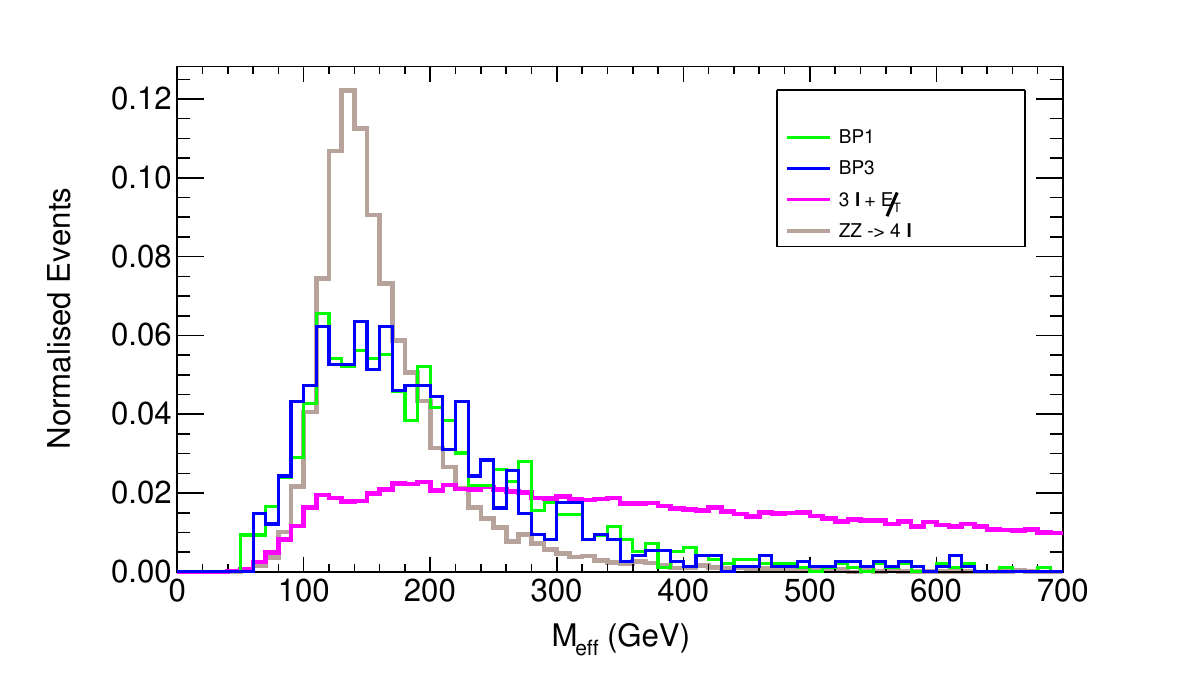}}
\subfigure[]{
\includegraphics[height = 6 cm, width = 8 cm]{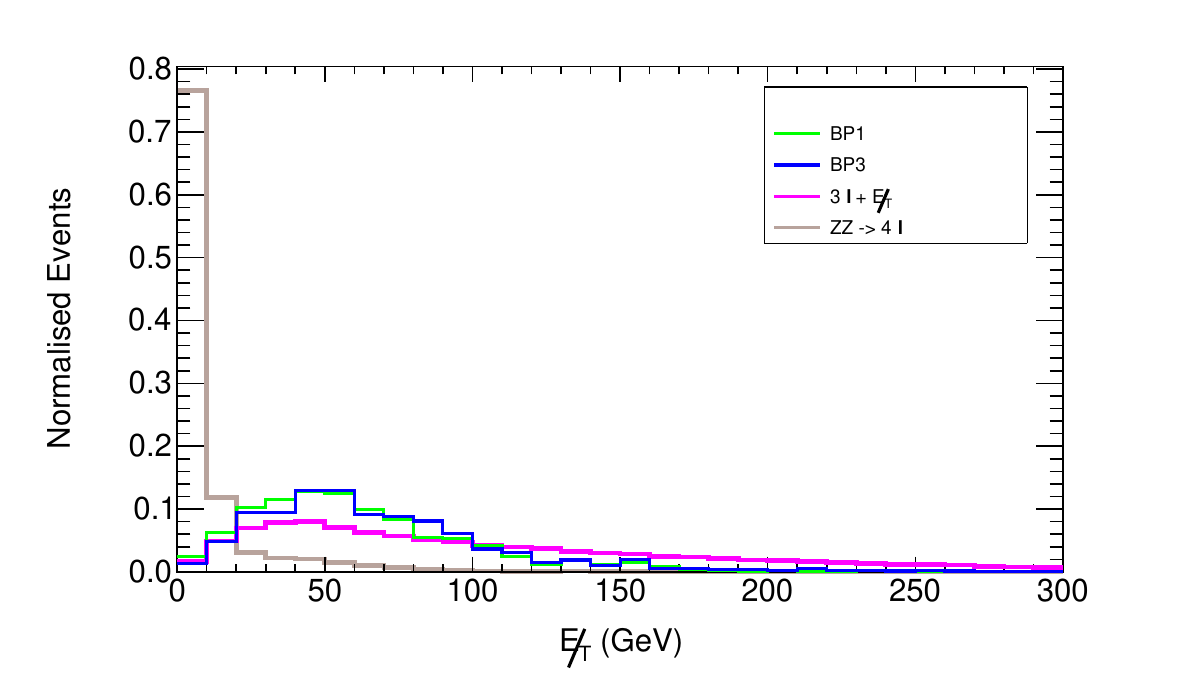}} \\
\subfigure[]{
\includegraphics[height = 6 cm, width = 8 cm]{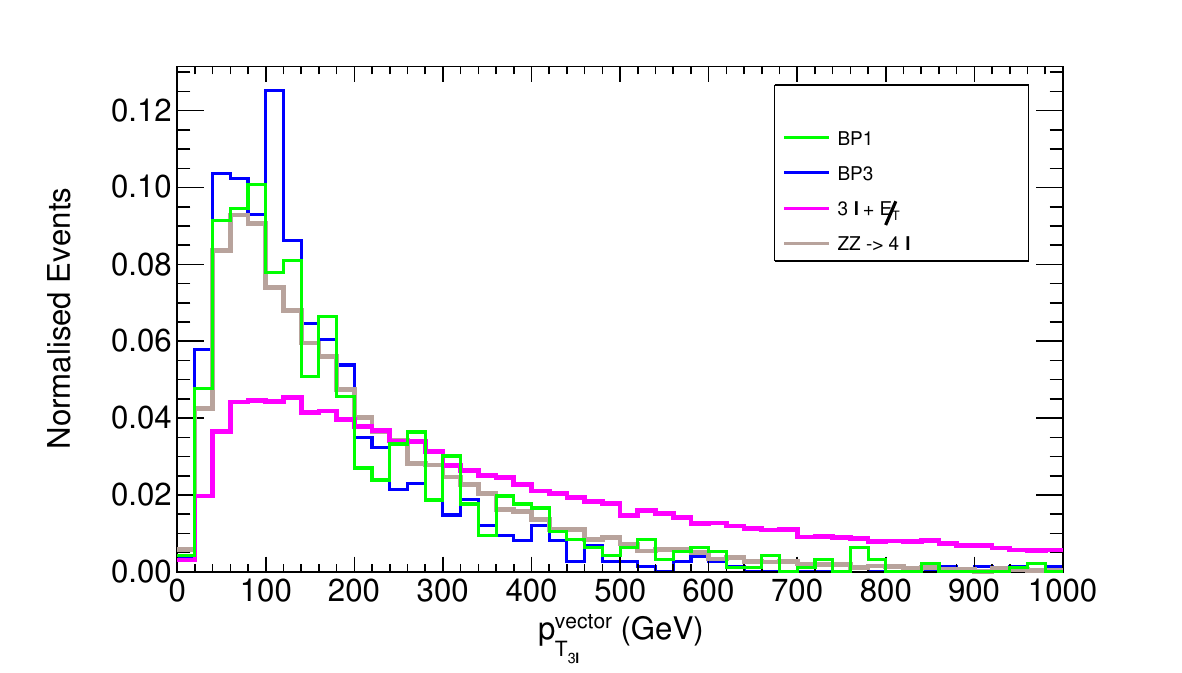}} 
\subfigure[]{
\includegraphics[height = 6 cm, width = 8 cm]{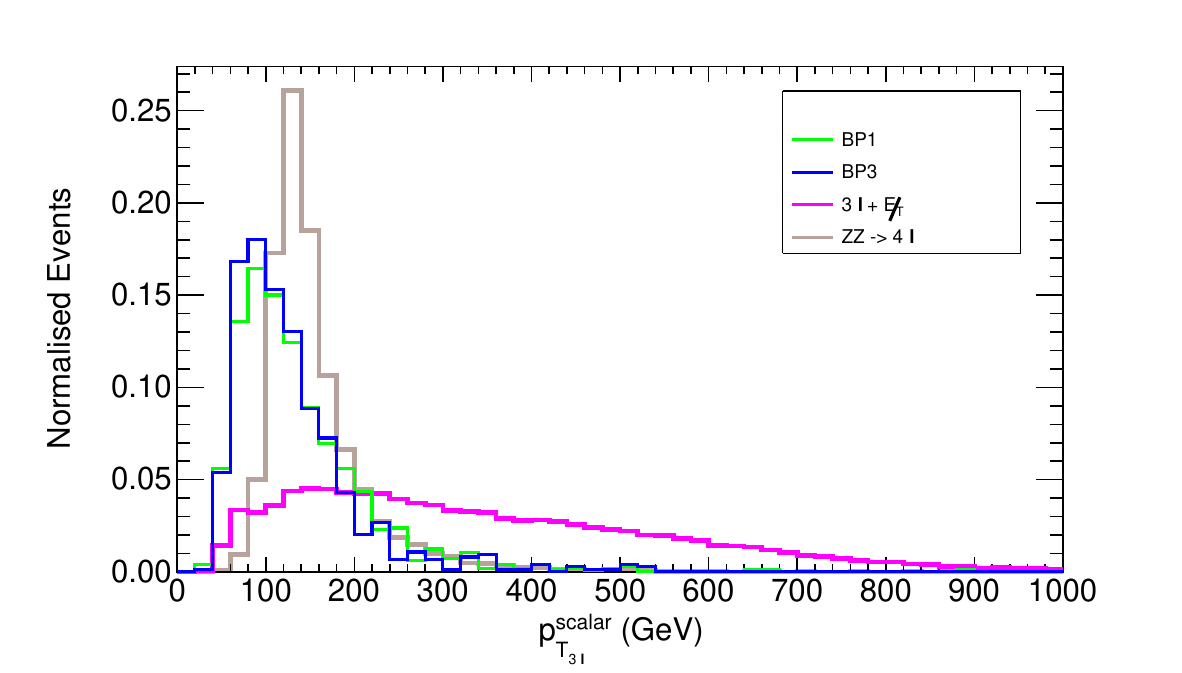}} \\
\subfigure[]{
\includegraphics[height = 6 cm, width = 8 cm]{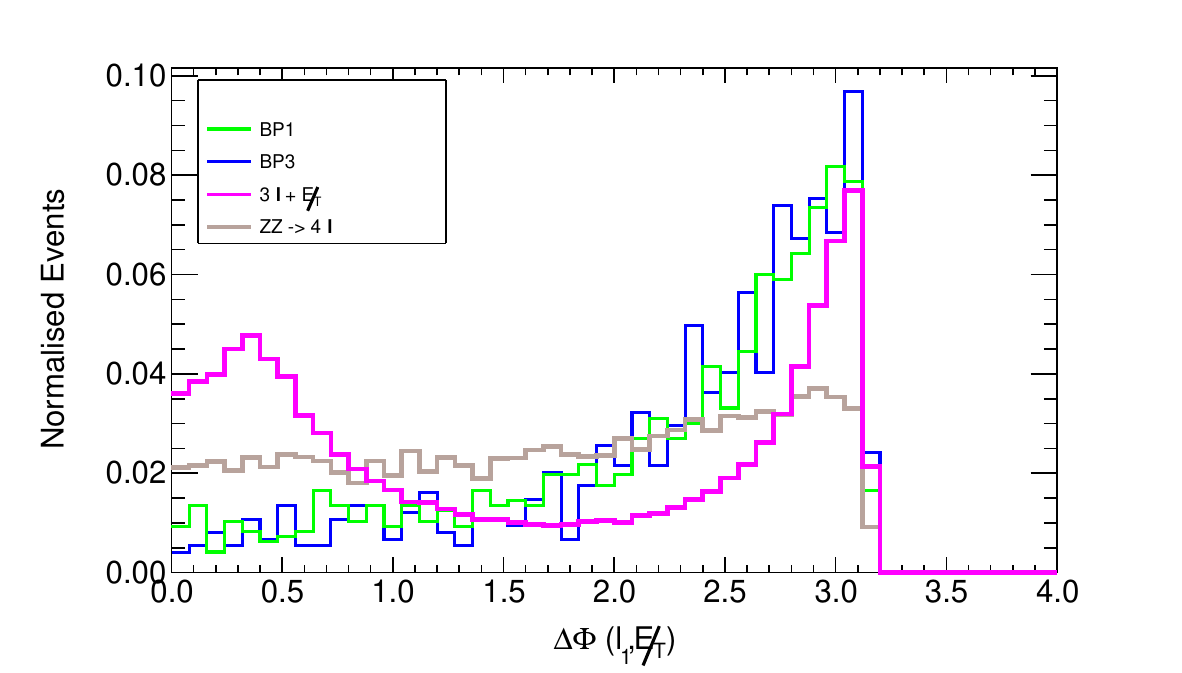}}} 
\caption{ Normalised distributions of $M_{\rm eff},~ \mET,~p_{T, tot}^{\rm vector}, ~ p_{T, tot}^{\rm scalar}, ~ \Delta \phi_{\ell_1, \mET}$ for $3 \ell + \mET$ channel at 14 TeV HL-LHC. }
\label{distribution-trilep}
\end{figure}
For this channel with more leptons, which is cleaner with smaller SM background, we restrict ourselves to the cut-based 
analysis only. To discriminate the signal from background, we demand our final state to have exactly three charged leptons 
with $p_T^\ell > 10$ GeV out of which two leptons are of the same sign and the third lepton is of opposite sign. 
Among these three leptons, at least two are expected to be of same flavour. We also impose $b$-jet veto 
(reject $p_T(b) > 20$ GeV) to eliminate the $b$-jets in the final state coming from the $t\bar{t}$ background. 
Next we identify a few kinematic variables which would help to discriminate the signal from background as follows: 
\begin{table}[ht!]
	\centering
		\resizebox{12cm}{!}{
	\begin{tabular}{|p{3.0cm}|c|c|c|c|c|p{3.0cm}|}
		\cline{2-6}
		\multicolumn{1}{c|}{}& \multicolumn{5}{|c|}{Number of Events after cuts ($\mathcal{L}=3$ ab$^{-1}$)} & \multicolumn{1}{c}{}  \\ \cline{1-6}
		SM-background  
		 & $C_1$  &  $ C_2 $    &  $C_3$  & $C_4$  &$C_5$ & \multicolumn{1}{c}{}
		\\ \cline{1-6} 
                  3$\ell + \mET$ & 129968  & 74974 & 64107 & 19530 &  13879   \\ \cline{1-6} 
		 $ZZ \to 4\ell$    & 6142  &  6117 &  520 & 308  & 270   \\ \hline \hline
		                     
			\multicolumn{1}{|c|}{Signal }  & \multicolumn{5}{|c|}{} & \multicolumn{1}{|c|}{Significance ($\mathcal{S}$)} \\ \cline{1-7} 
		\multicolumn{1}{|c|}{BP1}  & 2793   & 2732 & 2242 & 1376 & 1316 & \multicolumn{1}{|c|}{11.1} \\ \hline
		\multicolumn{1}{|c|}{BP2}  & 955  & 951  &  808 & 549  & 533 & \multicolumn{1}{|c|}{4.5} \\ \hline
		\multicolumn{1}{|c|}{BP3}   &  746  & 717   &  616 &  447 & 415 & \multicolumn{1}{|c|}{3.5} \\ \hline
		\multicolumn{1}{|c|}{BP4} &  324  & 310 & 263 & 160 & 147 & \multicolumn{1}{|c|}{1.2} \\ \hline 
		\multicolumn{1}{|c|}{BP5} &  117 & 114 & 94  & 61 & 28 & \multicolumn{1}{|c|}{0.5} \\ \hline 
		
	\end{tabular}}
      
	\caption{ The cut-flow for signal and backgrounds along with the significance for BP1, BP2, BP3, BP4 and BP5 at 14 TeV LHC for 3 ab$^{-1}$ integrated luminosity for the $ p p \rightarrow 3 \ell+\mET$ channel. }
	\label{tab:trilep_cut}
\end{table}	
\begin{itemize}
\item $C_1$ : Out of two same sign leptons and one opposite sign lepton in the final state, one can construct 
two invariant mass system $(M_{\ell^+ \ell^-})_{1,2}$,  considering one same and one opposite sign lepton 
at a time. Demanding $((M_{\ell^+ \ell^-})_{1,2} < 75$ GeV and $(M_{\ell^+ \ell^-})_{1,2} > 105$ GeV one can 
get rid of the $Z$-peak, which in turn reduces the $W^\pm Z$, $ZZ$ background drastically. We also impose a 
lower cut $(M_{\ell^+ \ell^-})_{1,2} > 12$ GeV to suppress the Drell-Yan background~\cite{Chatrchyan:2014aea}.
\item $C_2$ :  We define a variable $M_{\rm eff}$ as the scalar sum of all the lepton $p_T$'s and the $\mET$. 
In Fig.\ref{distribution-trilep}(a) the distribution of $3 \ell + \mET$ background (magenta line) is flatter and smeared 
with respect to the distributions of the signal (green and blue lines) and other background $ZZ$ (brown line). 
Setting $M_{\rm eff} < 500$ GeV helps in reducing the background. 
\item $C_3$ : Since the background $ZZ$ does not have $\mET$ in the final state explicitly, corresponding $\mET$  
distribution peaks at lower value than the signal as can be seen from Fig.\ref{distribution-trilep}(b). Thus a minimum 
cut of $\mET > 30$ GeV helps to reduce the $ZZ$ background drastically as can be found in Table~\ref{tab:trilep_cut}. 
\item $C_4$: We choose the vector sum of three leptons ($p_{T, 3 \ell}^{\rm vector}$) and the scalar sum of the same 
($p_{T, 3 \ell}^{\rm scalar}$) and show their distributions in Fig.\ref{distribution-trilep}(c) and \ref{distribution-trilep}(d) 
respectively. We find that kinematic selections of $p_{T, 3 \ell}^{\rm vector} < $ 200 GeV and 
$p_{T, 3 \ell}^{\rm scalar} <$ 250 GeV helps to reduce the $3\ell+ \mET$ background efficiently.

\item $C_5$ : We also construct the azimuthal angle between the unpaired third lepton out of total three leptons in 
the final state and $\mET$ as $\Delta \phi_{\ell, \mET}~$. Corresponding distributions are shown in 
Fig.\ref{distribution-trilep}(e). We find that the choice $\Delta \phi_{\ell, \mET}>$ 1.5 on the events help in eliminating 
the SM background further.
\end{itemize}

The number of events for signal and background, surviving after imposing the selection cuts on the aforementioned 
variables with 3 ab$^{-1}$ integrated luminosity are quoted along with the significances in Table~\ref{tab:trilep_cut}.  
For the five benchmarks BP1, BP2, BP3, BP4, BP5, using the cut-based analysis, the signal significances 
are 11.1, 4.5, 3.5, 1.2, 0.5 respectively. This is a substantial improvement over the previous two final state topologies 
considered earlier. In fact for BP1, $\mathcal{L} \sim $ 609 fb$^{-1}$ of integrated luminosity is enough to achieve a 5$\sigma$ 
significance in the tri-lepton channel.

\subsection{Four-lepton final state}
\label{fourlep}

In this section, we analyse the final state comprising of $4 \ell + \mET$.
The $4\ell+\mET$ final state for the signal can be obtained from the following processes:
\begin{equation}
 p p \to N_i N_i, N_i \to N_1 \ell^{+} \ell^{-}, ~~{\rm with} ~~i = 2,3,..8.
\end{equation}
 \begin{table}[htpb!]
\begin{center}\scalebox{0.95}{
\begin{tabular}{|c|c|c|}
\hline
Signal   & Process at LHC & LO cross-section $\sigma$  (fb) \\ \hline \hline
 BP1  & & 0.054 \\
 BP2  & & 0.033 \\
 BP3  & $p p \rightarrow 4\ell + \mET$& 0.02 \\
 BP4 &  & 0.008 \\
 BP5 & & 0.003 \\ \hline \hline
\end{tabular}}
\end{center}
\caption{The LO cross-sections for the signal $p p \rightarrow 4 \ell + \mET$ .}
\label{tab:bp_lhc:fourlep}
\end{table}
The events are generated using the same generation-level cuts and following the same method discussed in 
subsection~\ref{monolep}. The most dominant SM background~\cite{Aaboud:2018zeb} that gives rise to the 
similar final state is $VVV,(V= W^{\pm},Z$). The next irreducible background that follows the signal is $ZZ \to 4\ell$. In 
principle, $t\bar{t}Z$ can also mimic the signal, but putting a b-veto( rejecting $p_T(b) > 20$ GeV) kills the background.  
The other SM process $Z+2~jets$ also results in the same topology if the jets are mis-tagged as leptons. However, we 
find out that this background can be reduced considerably when proper $\mET$ cut is applied. Due to large 
cross-section, $t\bar{t}$ could also be a possible background. But demanding four lepton with $p_T^\ell > 10$ GeV 
and putting a $b$-jet veto reduces the background drastically. Thus from now on we shall only consider the 
dominant background $VVV$ and $ZZ \to 4\ell$ \footnote{The chosen benchmarks are also validated with an 
existing study by ATLAS at $\sqrt{s} = 13$ TeV \cite{Aaboud:2018zeb}.}. In Table~\ref{tab:bp_lhc:fourlep} we 
have tabulated the LO cross-sections for signal and background at 14 TeV LHC.

To disentangle the signal and background, we select four leptons with $p_T^\ell > 10$ GeV and 
$|\eta_\ell| < 2.5$ and reject any additional charged lepton satisfying the same. We also apply a veto on 
light-jets and $b-$jet in the final state. We consider the following set of kinematic variables to  
improve the the signal sensitivity over the background:

\begin{itemize}
 \item $D_1$: Out of total four leptons in the final state, we first select two pairs of leptons (pairwise of same flavour and 
 opposite sign), performing all possible combinations. Then we calculate the invariant mass of the pairs and compare whether 
 they are close to $M_Z$ or not. Considering the invariant mass of the first and second pair as 
 $(M_{\ell^+ \ell^-})_{1,2}$ respectively,  we reject all events where $ 105 > (M_{\ell^+ \ell^-})_{1,2} > 75$ GeV to exclude the 
 $Z$-peak of $ZZ$-background. For the signal, the four charged leptons are not produced from the decay of two $Z$-bosons,  
 which makes this cut very useful in boosting the signal significance.
 
 
 \item $D_2$: For the $ZZ$-background, the only source of $\mET$ is the mis-tagging of one or more leptons and so the 
 $\mET$  distribution for the dominant $ZZ$ background peaks at lower value than the signal which is expected as the 
 $\mET$ in $ZZ$ process mainly comes from mis-measurements. The $VVV$ background on the other hand would still 
 have a substantial overlap with the signal distribution which has a softer $\mET$ due to the compression in the 
 spectrum which leads to the cascade decays.
 Thus we choose a moderately low cut of $\mET > 30$ GeV which helps to reduce the $ZZ$ background significantly while 
 not killing too many of the signal events as we can see in Table~\ref{tab:fourlep_cut}. 
 
 \item $D_3$: We define the kinematic variable $M_{\rm eff}$ as the sum of the transverse momenta of four leptons and 
 $\mET$.  The background in this case seems to have a longish tail compared to the signal. To exclude the tail of the 
 $M_{\rm eff}$ distribution of the $ZZ$-background 
 we demand $M_{\rm eff} < 500$ GeV to enhance the signal significance.
\end{itemize}

The cuts applied on the aforementioned kinematic variables along with the significances are
listed in Table~\ref{tab:fourlep_cut}. 
For the five benchmarks, the significances at the integrated luminosity 3 ab$^{-1}$ are 10.1, 7.1, 5.2, 3.2, 1.0 respectively. 
Note that the first four benchmarks seem to achieve a significance $> 3 \sigma$ (the first three having $\mathcal{S} > 5\sigma$). Thus we find that the 
higher lepton multiplicity of the final states tend to achieve a more significant signal sensitivity in our model which is 
expected due to the addition of vector like fermions which decay to charged leptons.
\begin{table}[ht!]
	\centering
		\resizebox{12cm}{!}{
	\begin{tabular}{|p{3.0cm}|c|c|c|p{3.0cm}|}
		\cline{2-4}
		\multicolumn{1}{c|}{}& \multicolumn{3}{|c|}{Number of Events after cuts ($\mathcal{L}=3$ ab$^{-1}$)} &  \multicolumn{1}{c}{} \\ \cline{1-4}
		SM-background  
		 & $D_1$  &  $ D_2 $    &  $D_3$    & \multicolumn{1}{c}{}
		\\ \cline{1-4} 
           $VVV$ &  27 &  24 & 15   \\ \cline{1-4}
		 $ZZ \to 4 \ell$   & 5015 & 12 & 6  
		  \\ \hline \hline
		                      
			\multicolumn{1}{|c|}{Signal }  &\multicolumn{3}{|c|}{} &  \multicolumn{1}{|c|}{Significance ($\mathcal{S}$)} \\ \cline{1-5} 
		\multicolumn{1}{|c|}{BP1}    & 111  & 61 & 61 & \multicolumn{1}{|c|}{10.1}   \\ \hline
		\multicolumn{1}{|c|}{BP2}  & 64 & 40 & 40 & \multicolumn{1}{|c|}{7.1} \\ \hline
		\multicolumn{1}{|c|}{BP3}   & 32   &  28 & 28 & \multicolumn{1}{|c|}{5.2} \\ \hline
		\multicolumn{1}{|c|}{BP4} &  23  & 18 & 17 & \multicolumn{1}{|c|}{3.2} \\ \hline 
		\multicolumn{1}{|c|}{BP5} & 8  & 5 & 5 & \multicolumn{1}{|c|}{1.0} \\ \hline 
		
	\end{tabular}}

	\caption{ The cut-flow for signal and backgrounds along with the significances for BP1, BP2, BP3, BP4 and 
	BP5 at 14 TeV HL-LHC for 3 $ab^{-1}$ integrated luminosity for the $ p p \rightarrow 4 \ell+\mET$ channel. }
	\label{tab:fourlep_cut}
\end{table}	
\\

Before concluding this section, let us present a comparative study among all the aforementioned channels according 
to the degree of performance. For convenience, we have tabulated the signal significances corresponding to all BPs for all 
channels at an integrated luminosity 3 ab$^{-1}$ in Table \ref{compare}.  For   $1 \ell + \mET$ and 
$2 \ell + \mET$ channels we present the signal significances both for by the BDTD and cut-based analysis, whereas for 
$3 \ell + \mET$, $ 4 \ell + \mET$ channels we present the significances obtained from the cut-based analysis only. 
For all the benchmarks and all the channels there exist a generic pattern, which shows that the signal significance 
goes down with increasing DM masses. As mentioned earlier, smaller signal cross-section for larger DM masses 
are accountable for this particular pattern. According to the degree of performance, the $4 \ell + \mET$ channel 
fares the best among all for BP2, BP3, BP4 and BP5. With 3 ab$^{-1}$ luminosity, first four benchmark points can be probed with significance > $3 \sigma$.  Next better performing channel after $ 4 \ell + \mET$ is $3 \ell + \mET$ for last four benchmarks. In fact for BP1, $3 \ell + \mET$ channel turns out to be best performing with signal significance 11.1 at 3 ab$^{-1}$ integrated luminosity. For $1 \ell+ \mET$ and $2 \ell+ \mET$ channel the significance for BP1 are 7.0 and 6.7 respectively with cut-based analysis, which is improved to 7.8 using the BDTD analysis. 

  \begin{table}
 \begin{tabular}{|c|c|c|c|c|}
         \cline{2-5}
     \multicolumn{1}{c|}{}& \multicolumn{4}{|c|}{ Significance reach with 3 $ab^{-1}$ luminosity}\\  \cline{1-5}
   Benchmark points   & mono-lepton & di-lepton & tri-lepton & four-lepton \\ \cline{2-3} 
    & BDTD (Cut-based) & BDTD (Cut-based) & & \\ \cline{1-5}
  BP1 & 7.8 (7.0) & 7.8 (6.7)   & 11.1  & 10.1  \\ \hline
  BP2 & 5.9 (5.3) & 4.0 (2.5) & 4.5  & 7.1 \\ \hline
  BP3 & 2.6 (2.4) & 1.7 (1.3) & 3.5 & 5.2 \\ \hline
  BP4 & 0.5 (0.4) & 0.9 (0.8)  & 1.2  & 3.2 \\ \hline
  BP5 & 0.4 (0.3) & 0.32 (0.3)  & 0.5  & 1.0 \\ \hline
\hline
 \end{tabular}
 \caption{Significance reach with 3 $ab^{-1}$ luminosity for all the five benchmark points for mono-lepton, 
 di-lepton, tri-lepton and four lepton channel respectively.}
 \label{compare}
 \end{table}


\section{Conclusion}
\label{conc}
In this work, we extend the $S_3$-symmetric 2HDM with two generations of VLLs. The introduction of two 
generations of VLLs in the minimal version of the model is essential to ensure $S_3$-symmetric Yukawa Lagrangian. 
Since the VLLs are odd under the imposed $Z_2$-symmetry and the SM fermions are even under the same, 
the mixing between the SM leptons and the VLLs is forbidden. Thus we end up with a dark sector in our model which talks to 
the SM matter fields only through the SM force mediators and the scalar sector. In this set up, the lightest neutral VLL mass 
eigenstate serves as a viable DM candidate. 

Having satisfied the constraints like perturbativity, vacuum stability, electroweak precision data and Higgs signal 
strength, we show that a large portion of parameter space spanned by the model parameters is allowed from the observed 
relic density, direct and indirect detection experiments. We choose five representative points BP1, BP2, BP3, BP4 and BP5 
according to the low, medium and high DM masses to perform the collider analysis of some particular channels with 
$1 \ell + \mET, ~ 2\ell + \mET, ~ 3\ell + \mET$ and $4 \ell +\mET$ in the final state at 14 TeV HL-LHC. We must point 
out here that our choice of 
benchmarks for the LHC analysis only represents a part of the parameter space allowed by DM data, as it gives a 
relatively lighter spectrum which can give appreciable signal sensitivity to our model at LHC. To highlight this, we have  
chosen a benchmark point (BP5) which represents a point near the threshold (for the spectrum) which will be out 
of reach at LHC, even with the very high luminosity (vHL-LHC) option.  
  
To start with, we first analyse the final state containing $1\ell + \mET$, which can originate from the pair production 
of the charged VLLs and neutral VLLs as well as from the associated production of the charged and neutral VLLs. The major 
background for this channel is $p p \to W^\pm \to 1 \ell^{\pm} + \mET$.  With traditional cut-based analysis, we show that 
with 3 ab$^{-1}$ luminosity BP1 and BP2 can be probed with significance $> 5\sigma$, which in turn improves with 
multivariate (BDTD) analysis. Next we move on to perform the collider analysis of the final state comprising of 
$2\ell + \mET$ (di-lepton along with missing transverse energy), which mainly comes from the pair production of the 
charged VLLs and neutral VLLs individually. The main background is $p p \to 2\ell + \mET$ which takes care of 
$W^{+}W^{-}, ZZ$ pair production. $t\bar{t}$ and $W^{\pm}Z$ also contribute as subdominant SM background for the 
di-lepton channel. After performing cut-based analysis we find that only BP1 can be probed with a significance 
$> 5 \sigma$ while the sensitivity in the di-lepton channel diminishes for the rest of the benchmarks. The primary reason 
for this is the fact that the mass splittings amongst the VLL's is not too large which leads to a relatively compressed spectrum. 
The resulting decay products in the cascade are therefore not very hard, leading to a significant overlap of the kinematic 
distributions with that of the SM background. This led to 
less signal significance in the di-lepton mode, though it can be improved marginally with a BDTD analysis. 
The final state containing tri-lepton with missing transverse energy can be generated from the associated production of the 
charged and neutral VLLs. Corresponding irreducible background originates from $p p \to 3\ell + \mET$ process. The situation 
is found to improve here as the SM background is now smaller compared to the di-lepton final state. With simple 
cut-based analysis we find out that with 3 ab$^{-1}$ luminosity,  BP1 can be probed with significance $> 10\sigma$, while 
now even BP3 has a $>3\sigma$ sensitivity. The situation improves further with an increase in charged lepton multiplicity, 
which we show with the analysis of $4\ell + \mET$ final state, arising mostly from neutral VLL pair production. The major SM 
background for this process is $ p p \to ZZ \to 4\ell$ and $p p \to V\,V\,V \,\, (V\equiv W,Z)$ which have small cross-sections. 
We find that now the four benchmark points can be probed with significance > 3 $\sigma$ with 3 ab$^{-1}$ 
integrated luminosity. Thus comparing all four channels of the multi-lepton final states we find that $4 \ell + \mET$ 
channel turns out to be the most promising among all owing to it being the most clean and background free final state. 

We conclude our discussion by stating that this model can provide a viable Majorana type DM candidate and that a part of the 
allowed parameter space (with DM masses up to $\sim 300$ GeV) can be tested at the 14 TeV HL-LHC in the 
multi-lepton channel. The relative compression in the mass spectrum  of the VLL's do not allow very clean kinematic 
thresholds that could provide as a good discriminator for signal  against the SM background. This limits the search 
sensitivity of the model to relatively light VLL masses of about 350 GeV, beyond which it is very difficult to achieve any signal 
sensitivity even with the HL-LHC option. To probe higher DM masses one may benefit by looking for such a model at 
the 1 TeV ILC which warrants a separate study in future \cite{S3atILC}. 


\section{Acknowledgement}
IC acknowledges support from DST, India, under grant number IFA18-PH214 (INSPIRE Faculty Award). NG and SKR would 
like to acknowledge support from the Department of Atomic Energy, Government of India, for the Regional Centre for 
Accelerator-based Particle Physics (RECAPP). NG and SKR would like to thank Tousik Samui and Anjan Barik for 
computational help.

\appendix

\section{\\ Decay width of $h \rightarrow \gamma \gamma$}
\label{app : A}
Amplitude and decay width of the process $h \rightarrow \gamma \gamma$ can be written as \cite{Djouadi:2005gj}:
\besub
\bea
\mathcal{M}_{h \to \gamma \gamma} &=& 
\sum_f N_f Q_f^2 f_{hff} A_{1/2}\Big(\frac{M^2_h}{4 M^2_f}\Big)
 + f_{h VV} A_1\Big(\frac{M^2_h}{4 M^2_W}\Big)\nonumber \\
&& 
 +  \frac{\l_{h H^+ H^-} v}{2 M^2_{H^+}} A_0\Big(\frac{M^2_h}{4 M^2_{H^+}}\Big) + \sum_{i=1}^4 \frac{\lambda_{h E_i^+ E_i^-} v}{M_{E_i}} A_{1/2}\Big(\frac{M_h^2}{4 M_{E_i^2}}\Big)\\ 
\Gamma_{h \to \gamma \gamma} &=& \frac{G_F \a^2 M_h^3}{128 \sqrt{2} \pi^3} |\mathcal{M}_{h \to \gamma \gamma}|^2,
\eea
\eesub
where $N_f, Q_f, G_F$ and $\a$ are respectively color factor, charge of fermion, the Fermi constant and the QED fine-structure constant. For quarks $N_f = 3$. $\lambda_{h H^+ H^-}$ and $\lambda_{h E_i^+ E_i^-} ~(i=1,2,3,4)$ are $h H^+ H^-$ and $h E_i^+ E_i^-$ couplings respectively. $f_{hff}, f_{hVV}$ are scale factors of $hff, hVV$ couplings with respect to SM. At exact alignment limit,
\bea
f_{hff} = f_{hVV} = 1
\eea

The loop functions can be written as,
\besub
\bea
A_{1/2}(x) &=& \frac{2}{x^2}\big((x + (x -1)f(x)\big), \\
A_1(x) &=& -\frac{1}{x^2}\big((2 x^2 + 3 x + 3(2 x -1)f(x)\big), \\
A_0(x) &=& -\frac{1}{x^2}\big(x - f(x)\big),  \\
\text{with} ~~f(x) &=& \text{arcsin}^2(\sqrt{x}); ~~~x \leq 1 
\nonumber \\
&&
= -\frac{1}{4}\Bigg[\text{log}\frac{1+\sqrt{1 - x^{-1}}}{1-\sqrt{1 - x^{-1}}} -i\pi\Bigg]^2; ~~~x > 1.
\eea
\eesub

where $A_{1/2}(x), A_1(x)$ and 
$A_0(x)$ are the respective amplitudes for the spin-$\frac{1}{2}$, spin-1 and spin-0 particles in the loop respectively while $\lambda_{h H^+ H^-} = 2 v~ (\lambda_1 - \lambda_3)$.  
\section{Decay modes of the charged and neutral VLLs}
\label{app : B}
We list below the tables which summarise the decay probabilities of the VLLs, both charged and neutral for the five benchmark
points chosen for our analysis.
\begin{table}[htbp!]
	\centering
	\resizebox{17cm}{!}{
	\begin{tabular}{|l|c|c|c|c|c|}
		\hline
Benchmark Points &  $E^{\pm}_1$ decay  & $E^{\pm}_2$ decay & $E^{\pm}_3$ decay  & $E^{\pm}_4$ decay  \\ \hline
\hspace{5 mm} BP1 
& 63.5\% $W^{\pm} N_1$,& 46.7\% $ W^{\pm} N_2$, 35.0\% $ W^{\pm} N_1$, & 33.7\% $ W^{\pm} N_1$, 26.4\% $ W^{\pm} N_4$, & 60.8\% $ W^{\pm} N_1$,  \\
& 36.5\% $ W^{\pm} N_2$ & 10.3\% $ W^{\pm} N_4$, 7.9\% $ W^{\pm} N_3$ & 18.3\% $ W^{\pm} N_2$,  & 39.2\% $ W^{\pm} N_2$\\ 
     & & & 12.7(1.28)\% $ E^{\pm}_1 h(Z)$,&    \\ 
     & & & 3.97\% $ W^{\pm} N_3$, 3.39\% $ W^{\pm} N_5$  &    \\ \hline
\hspace{5 mm} BP2 & 94.6\% $ W^{\pm} N_1$, & 39.7\% $ W^{\pm} N_1$, 35.6\% $ W^{\pm} N_2$, & 36.3\% $ W^{\pm} N_3$,
35.4\% $ W^{\pm} N_1$, & 53.0\% $ W^{\pm} N_2$, 40.7\% $ W^{\pm} N_1$, \\
    & 5.4\% $ W^{\pm} N_2$ & 21.5\% $ W^{\pm} N_3$, 3.1\% $ W^{\pm} N_4$ 
    & 20.7\% $ W^{\pm} N_2$, 3.0\% $ W^{\pm} N_4$, & 6.3\% $ W^{\pm} N_3$  \\ 
    & &  & 2.5\% $ W^{\pm} N_5$, 2.1\% $ E^{\pm}_1 Z$ &  \\ \hline
\hspace{5 mm} BP3 & 89.4\% $ W^{\pm} N_1$, & 60.0(1.1)\% $E^{\pm}_1 h(Z)$, & 30.8(1.8)\% $ E^{\pm}_1 h(Z)$,
& 53.3\% $ W^{\pm} N_1$,  \\
    & 10.6\% $ W^{\pm} N_2$ & 14.1\% $ W^{\pm} N_2$, 10.8\% $ W^{\pm} N_1$, & 16.9\% $ W^{\pm} N_1$, 14.6\% $ W^{\pm} N_2$,
    &46.7\% $ W^{\pm} N_2$ \\
    &  & 4.9\% $ W^{\pm} N_4$, 3.5\% $ W^{\pm} N_3$,&13.4\% $ W^{\pm} N_4$, 6.6\% $ W^{\pm} N_6$, &  \\
    &  & 3.4\% $ W^{\pm} N_6$, 2.2\% $ W^{\pm} N_5$ & 6.2\% $ W^{\pm} N_5$, 4.4\% $E^{\pm}_2 h$, 3.6\% $ W^{\pm} N_3$ &   \\ \hline 
\hspace{5 mm} BP4 & 100.0\% $ W^{\pm} N_1$ & 62.0\% $ W^{\pm} N_1$,24.5\% $ W^{\pm} N_5$,& 94.9\% $ W^{\pm} N_6$, &  99.3\% $ W^{\pm} N_2$,   
\\
 & & 11.6\% $ W^{\pm} N_3$, 1.4\% $ W^{\pm} N_2$ & 3.3\% $ W^{\pm} N_1$ &0.7\% $ W^{\pm} N_1$ \\ \hline
 \hspace{5 mm} BP5 & 100.0\% $ W^{\pm} N_1$ & 30.1\% $ W^{\pm} N_1$, 19.6\% $ W^{\pm} N_4$,& 56.4\% $ W^{\pm} N_1$, 17.3\% $ W^{\pm} N_2$, &100.0\% $ W^{\pm} N_1$     \\
 & & 16.7\% $ W^{\pm} N_3$, 14.4\% $ W^{\pm} N_2$,&14.7\% $ W^{\pm} N_4$, 8.0\% $ W^{\pm} N_3$, &  \\
 & & 11.2\% $ W^{\pm} N_5$, 7.9\% $ W^{\pm} N_6$& 2.4\% $ W^{\pm} N_6$, 1.2\% $ E^{\pm}_1 Z$ &    \\ \hline
\end{tabular}}
\caption{ Decay modes for charged VLL.}
\label{BR:charged}
\end{table} 

\begin{table}[h!]
	\centering
	\resizebox{14cm}{!}{
	\begin{tabular}{|l|c|c|c|c|c|}
		\hline
 Benchmark Points &  $N_2$ decay  & $N_3$ decay & $N_4$ decay  & $N_5$ decay  \\ \hline
 \hspace{5 mm} BP1  & 49.5\% $ \ell\bar{\ell} N_1$, & 68.5\% $ q\bar{q} N_1$,& 70.2(23.9)\% $ N_2 h(Z)$, & 81.8\% $ N_2 Z$, \\ 
         & 49.0\% $ q\bar{q} N_1$  & 31.5\% $ \ell\bar{\ell} N_1$ &  2.6\% $ N_4 h$, 2.2\% $ E^{\pm}_1 W^{\mp}$ & 18.2\% $ N_1 h$ \\ \hline
 \hspace{5 mm} BP2  & 84\% $ q\bar{q} N_1$, & 58.5\% $ q\bar{q} N_1$,& 67.8\% $ q\bar{q} N_1$, & 64.6(23.7)\% $ N_2 h(Z)$,  \\
        & 15.1\% $ \ell\bar{\ell} N_1$ & 41.5\% $ \ell\bar{\ell} N_1$ & 30.9\% $ \ell\bar{\ell} N_1$  & 10.7\% $ N_3 h$ \\
     \hline
\hspace{5 mm} BP3   & 47.0\% $ q\bar{q} N_1$, & 68.8\% $ q\bar{q} N_1$,  & 68.8\% $ q\bar{q} N_1$, & 67.7(4.9)\% $ N_2 Z(h)$   \\
          & 53.0\% $ \ell\bar{\ell} N_1$ & 31.2\% $ \ell\bar{\ell} N_1$  & 31.2\% $ \ell\bar{\ell} N_1$ & 27.4\% $ N_1 h$\\
     \hline
 \hspace{5 mm}    BP4  & 89.1\% $ q\bar{q} N_1$,   & 68.4\% $ q\bar{q} N_1$, & 68.3\% $ q\bar{q} N_2$, & 67.9\% $ q\bar{q} N_2$,  \\
     & 10.9\% $ \ell\bar{\ell} N_1$ & 31.6\% $ \ell\bar{\ell} N_1$ & 31.7\% $ \ell\bar{\ell} N_2$ & 32.1\% $ \ell\bar{\ell} N_2$  \\ \hline 
 \hspace{5 mm}   BP5    & 68.6\% $ q\bar{q} N_1$,  & 95\% $ q\bar{q} N_1$, & 83.1\% $ q\bar{q} N_1$, 4.3\% $ \ell\bar{\ell} N_1$, & 67\% $ q\bar{q} N_3$,31.2\% $ \ell\bar{\ell} N_3$ \\
       & 31.4\% $ \ell\bar{\ell} N_1$  & 5\% $ \ell\bar{\ell} N_1$ & 7\% $ q\bar{q} N_3$, 6.4\% $ \ell\bar{\ell} N_3$ & 1.6\% $ q\bar{q} N_1$   \\
     \hline
     \end{tabular}} 
      \caption{ Decay modes for neutral VLL. }
    \label{BR:neutral1}
     \end{table}
     
     \begin{table}[htbp!]
 	\centering
 	\resizebox{15.0cm}{!}{
 	\begin{tabular}{|l|c|c|c|}
 		\hline
 Benchmark Points & $N_6$ decay & $N_7$ decay & $N_8$ decay \\ \hline
 BP1  & 68.8\% $ q\bar{q} N_1$,  &  32(26)(18)\% $E^{\pm}_{1(2)(3)} W^{\mp}$, & 16.5\% $N_7 h$, 14.9(6.2)\% $ N_4 h(Z)$,\\ 
   & 31.2\% $ \ell\bar{\ell} N_1$ & 11.2\% $ N_2 Z$, 6.9\% $ N_6 Z$,   & 12.4(6.1)\% $ N_1 h(Z)$, 5.0\% $ N_1 H$,\\ 
    & & 2.4\% $ N_4 h$, 1.9\% $ N_1 h$ & 8.4(8.3)(5.2)\% $E^{\pm}_{2(1)(4)} W^{\mp}$, \\ 
   &  & & 3.9(1.5)\% $ N_6 h(H)$, 3.2(1.6)\% $ N_2 h(H)$ \\ \hline
 BP2  & 100\% $ N_2 Z$ & 32.3(27.2)(9.6)\% $ E^{\pm}_{1(2)(3)} W^{\mp}$, & 18.3(8.3)\% $ N_3 h(Z)$, 15.6(8.0)\% $ N_1 h(Z)$, \\
    & & 13.4\% $ N_2 Z$, 8.2\% $ N_6 Z$, & 10.7(10.6)(5.6)\% $E^{\pm}_{2(1)(4)} W^{\mp}$,\\
    & & 4.2\% $ N_3 h$, 3.2\% $ N_1 h$, & 7.3\% $ N_6 h$, 6.3\% $ N_2 h$,\\
    & & 4.6\% $ N_1 H$ & 3.4\% $ N_3 H$, 3.4\% $ N_1 H$,\\
    & & & 1.6\% $ N_2 H$ \\ \hline
BP3    & 70.4(27.9)\% $ N_2 h(Z)$, & 40.4(29.2)(2)\% $ E^{\pm}_{1(2)(3)} W^{\mp}$, & 22.9(8.9)\% $ N_4 h(Z)$, 18.5(9.7)\% $ N_1 h(Z)$,\\
        & 1.6 $N_1 Z$ & 13.2\% $ N_2 Z$, 8.2\% $ N_6 Z$, & 13.6(12)(1.3)\% $E^{\pm}_{1(2)(4)} W^{\mp}$, \\
    &  & 3.4\% $ N_4 h$, 2.7\% $ N_1 h$ & 5.1\% $ N_6 h$, 4.6\% $ N_2 h$, 2.0\% $ N_1 H$ \\ \hline
     
BP4   & 68.9\% $ q\bar{q} N_1$, & 59.3(7)\% $ N_5 Z(h)$, & 73.1(14.9)\% $ N_6 h(Z)$,\\
     & 31.1\% $ \ell\bar{\ell} N_1$  & 31.6\% $ E^{\pm}_3 W^{\mp}$  & 9.6(2)\% $ E^{\pm}_{2,(4)} W^{\mp}$ \\ \hline 
    
BP5  & 54.3\% $ q\bar{q} N_3$, 24.6\% $ \ell\bar{\ell} N_3$, & 79.1\% $ N_4 h$, 10.5(1.5)\% $ N_1 Z(h)$, & 26.8(21.1)\% $ N_1 h(Z)$,\\
    & 11.2\% $ q\bar{q} N_4$, 6.3\% $ \ell\bar{\ell} N_3$ & 5.8\% $ N_2 Z$, 2.2\% $ E^{\pm}_1 W^{\mp}$ & 21.8(14.7)\% $ N_2 h(Z)$, \\
     & & & 5.7\% $ N_3 h$, 5.8(4.1)\% $ E^{\pm}_{1(2)} W^{\mp}$\\ \hline
 \end{tabular}}
 \caption{ Decay modes for neutral VLL (cont.d).}
 \label{BR:neutral2}
 \end{table}
%
\bibliographystyle{JHEP}
 \bibliography{draft_S3_v2.bib}
\end{document}